\documentclass[aoas]{imsart}

\RequirePackage{amsthm,amsmath,amsfonts,amssymb}
\RequirePackage[authoryear]{natbib}
\RequirePackage[colorlinks,citecolor=blue,urlcolor=blue]{hyperref}
\RequirePackage{graphicx}
\usepackage{float}
\usepackage{booktabs}
\usepackage{multirow}
\usepackage{afterpage}
\usepackage{tikz}
\usepackage{verbatim}
\usepackage{enumitem}
\usepackage{bbm}
\usepackage{fix-cm}
\usepackage{xr}

\DeclareFontShape{T1}{ptm}{m}{scit}{<->ssub * ptm/m/it}{}

\startlocaldefs

\theoremstyle{plain}

\newtheorem{theorem}{Theorem}

\theoremstyle{remark}

\endlocaldefs

\begin{document}

\begin{frontmatter}
\title{A Data Envelopment Analysis Approach for Assessing Fairness in Resource Allocation: Application to Kidney Exchange Programs}
\runtitle{Fairness in Resource Allocation}

\begin{aug}
\author[A]{\fnms{Ali}~\snm{Kaazempur-Mofrad}\ead[label=e1]{amofrad@ucla.edu}\orcid{0000-0003-0537-9577}} \and 
\author[A]{\fnms{Xiaowu}~\snm{Dai}\ead[label=e2]{daix@ucla.edu}\orcid{0000-0002-5889-5201}\thanksref{t1}}
\address[A]{Department of Statistics and Data Science, University of California, Los Angeles\\ 
\printead{e1} \
\printead{e2}}
\thankstext{t1}{Address for correspondence: Xiaowu Dai, Department of Statistics and Data Science, University of California, Los Angeles, 8125 Math Sciences Bldg \#951554, Los Angeles, CA 90095, USA. Email: daix@ucla.edu.}
\end{aug}

\vspace{0.2in}
To appear in \emph{Annals of Applied Statistics}.

\vspace{0.1in}

\begin{abstract}
Kidney exchange programs have substantially increased transplantation rates but also raise critical concerns about fairness in organ allocation. We propose a novel framework leveraging Data Envelopment Analysis (DEA) to evaluate multiple dimensions of fairness—Priority, Access, and Outcome—within a unified model. This approach captures complexities often missed in single-metric analyses. Using data from the United Network for Organ Sharing, we separately quantify fairness across these dimensions: Priority fairness through waitlist durations, Access fairness via the Living Kidney Donor Profile Index (LKDPI) scores, and Outcome fairness based on graft lifespan. 
We then apply our conditional DEA model with covariate adjustment to demonstrate significant disparities in kidney allocation efficiency across ethnic groups. To quantify uncertainty, we employ conformal prediction within a novel Reference Frontier Mapping (RFM) framework, yielding group-conditional prediction intervals with finite-sample coverage guarantees. Our findings show notable differences in efficiency distributions between ethnic groups. Our study provides a rigorous framework for evaluating fairness in complex resource allocation systems with resource scarcity and mutual compatibility constraints. 
\end{abstract}


\end{frontmatter}

\section{Introduction}

Chronic kidney disease (CKD) represents a significant and growing public health challenge in the United States and globally. According to the \citet{CDC2023}, CKD affects approximately 14\% of U.S. adults, or 35.5 million people. This prevalence has been steadily increasing over the past decades, driven by rising rates of diabetes, hypertension, and an aging population \citep{Coresh2007}. Of particular concern is the subset of CKD patients who progress to end-stage renal disease (ESRD), a life-threatening condition where the kidneys have lost over 85\% of their function. In 2020, over 805,000 Americans were living with ESRD, with 130,719 new cases diagnosed that year \citep{USRDS2024}. For these patients, kidney transplantation offers the best outcomes in terms of quality of life, long-term survival, and cost-effectiveness compared to dialysis \citep{Wolfe1999, Axelrod2018}. However, the stark reality is that the demand for donor kidneys far exceeds the available supply. For example, in 2020, while 23,642 kidney transplants were performed in the U.S., nearly 100,000 patients remained on the waiting list \citep{Lentine2022}. This critical shortage has spurred innovations in transplantation strategies, with kidney exchange programs emerging as a particularly promising solution. These programs, first proposed by \citet{Rapaport1986} and later formalized by \citet{Roth2004}, allow patients with willing but incompatible living donors to exchange donors with other incompatible pairs. Kidney paired donation (KPD) programs have significantly increased the number of living donor transplants performed annually in the United States \citep{Massie2013}, offering new hope to patients who might otherwise face prolonged waiting times or potential ineligibility for transplantation. These programs have been particularly beneficial for hard-to-match patients, including those with high panel reactive antibody levels or blood type O recipients \citep{Gentry2011}. Moreover, recent advancements, such as global kidney exchange \citep{Rees2017} and the incorporation of compatible pairs \citep{Gentry2007}, have further enhanced the potential of these programs, thereby expanding the pool of potential donors and recipients. 

While kidney exchange programs have undoubtedly improved overall transplantation rates, they raise pressing questions about fairness in organ allocation. These programs, typically designed to maximize the total number of transplants, may inadvertently disadvantage certain subgroups of the population \citep{Agarwal2019}. Examining fairness in such systems poses significant challenges, particularly when multiple, potentially conflicting criteria must be considered simultaneously. Throughout this paper, we focus on group fairness, which aims for parity across demographic groups \citep{Hardt2016}. This concept is particularly relevant in kidney allocation, where disparities in access and outcomes among different ethnic groups have been well-documented. For instance, \citet{Malek2011} found that only 57\% of Asian patients received a transplant within five years compared to nearly 70\% of White patients, 
while \citet{Fan2010} found that Asian patients experience the highest graft survival rates among all ethnic groups. Such conflicting outcomes across different metrics underscore the need for a comprehensive framework to evaluate and enhance fairness in kidney allocation systems.

We propose a general framework for evaluating fairness in KPD allocation using conditional Data Envelopment Analysis (DEA). Originally developed by \citet{Charnes1978} for measuring the relative efficiency of decision-making units, DEA offers a flexible approach for evaluating multiple inputs and outputs without requiring a priori specification of their relative importance. Our framework employs a hyperbolic graph efficiency DEA model \citep{Fare1985} extended to account for exogenous covariates via conditional frontiers \citep{buadin2012measure}. Prior to applying this conditional DEA framework, we conduct individual analyses of three key fairness components—Priority, Access, and Outcome Fairness—using specialized models to show complex and sometimes conflicting patterns of disparity across ethnic groups. For Priority Fairness, we employ mediation analysis, aligned with the work of \citet{Imai2010}, to disentangle the direct and indirect effects of patient characteristics on waiting times. Access Fairness is evaluated through counterfactual analysis of LKDPI score disparities \citep{Kusner2017}. Outcome Fairness evaluates graft survival through competing risks analysis \citep{PintoRamirez2022} to examine graft rejection risks in the presence of other competing causes of graft failure. Our analysis shows that while Asian patients face longer waitlist times, they demonstrate better graft survival outcomes. Conversely, White patients enjoy shorter waitlist durations but face higher risks of graft failure due to rejection. Black patients, while experiencing moderate waitlist times, face the highest risk of graft rejection. These conflicting results highlight the challenge of achieving fairness across all criteria simultaneously in kidney allocation. Our conditional DEA approach addresses these complexities by minimizing unfavorable inputs (e.g., long waiting times) while maximizing desirable outputs (e.g., graft survival outcomes). The resulting hyperbolic efficiency measure provides a single score that captures how well KPD allocations serve each patient or subgroup, considering multiple fairness criteria simultaneously.

To quantify uncertainty around these efficiency scores, we integrate conformal prediction techniques proposed by \citet{Gibbs2023} within the conditional DEA framework, using our proposed Reference Frontier Mapping (RFM) procedure to decouple frontier estimation from evaluation. This procedure generates group-conditional prediction intervals for efficiency scores with finite-sample coverage guarantees, ensuring equitable uncertainty quantification across ethnic groups. Our approach provides a new method for evaluating fairness in kidney allocation and offers several key contributions to the field. By mapping efficiency scores across demographic groups and comparing their distributions, we identify potential disparities and areas for improvement in the allocation system, which may not be apparent when considering individual fairness criteria in isolation. The comprehensive efficiency score provides an interpretable, unified measure of overall system fairness for each patient or subgroup, and integrating uncertainty quantification enhances the reliability of our findings. 

Our fairness framework for kidney allocation expands on existing methods, which often rely on single metrics, such as mechanism design models \citep{Ogryczak2014, Chorppath2011}, multi-criteria decision analysis with predetermined weights \citep{MorenoCalderon2020, Thokala2016}, or specific survival comparisons \citep{Cohen2015, Tonelli2011}. The proposed framework provides a unified efficiency measure that integrates multiple fairness criteria without requiring fixed weights or distributional assumptions. This approach can be applied more broadly to other resource allocation systems facing scarcity and compatibility constraints, such as healthcare \citep{Chilingerian1995, Ozcan2014} and education \citep{Shero2022}, where fairness is critical.

The remainder of this paper is structured as follows. Section~\ref{sec:2} introduces the methods, including the conditional DEA model and RFM-based uncertainty quantification. Section~\ref{sec:3} details the data preparation process and the individual fairness analyses. Section~\ref{sec:4} presents the results of applying our conditional DEA framework to our data, along with statistical testing and inference. Section~\ref{sec:5} concludes the paper with potential future research.

\section{Methodology}
\label{sec:2}

\subsection{Fairness Criteria in Kidney Allocation}
\label{sec:2.1} 
Evaluating fairness in kidney allocation systems requires multiple criteria to capture the complexities of the transplantation process \citep{Persad2009}. Our analysis focuses on three key criteria: Priority Fairness, Access Fairness, and Outcome Fairness, with each targeting a critical aspect of the allocation system.

\subsection*{Priority Fairness}
Priority Fairness reflects the fairness in waitlist duration across different ethnic groups, focusing on how fairly patients are prioritized for transplantation. In our model, let $X_{1i}$ denote the waitlist duration (in days) for patient $i$. The goal is to evaluate disparities in $X_1$ across demographic groups, with shorter durations indicating higher priority. Studies by \citet{Wolfe1999} and \citet{Gill2005} emphasize the crucial impact of waitlist duration on fair and timely access to transplantation. We conduct a mediation analysis in Section~\ref{sec:3.3} to examine the direct effect of ethnicity on $X_{1i}$ while accounting for mediators, such as recipient blood type and dialysis status.

\subsection*{Access Fairness}

Access Fairness considers the distribution of kidney quality, as indicated by the Living Kidney Donor Profile Index (LKDPI). The LKDPI score is a numerical measure that combines both donor and recipient characteristics to summarize the relative risk associated with a living donor transplant \citep{massie2016risk}. The LKDPI can theoretically take values below 0 (indicating lower risk than any deceased donor kidney) or above 100 (indicating higher risk than any deceased donor kidney). In our model, let $X_{2i}$ denote the LKDPI score for the kidney allocated to patient $i$. By studying $X_2$ across ethnic groups, this criterion suggests whether there are disparities in the quality of organs distributed. In Section~\ref{sec:3.3}, we conduct a counterfactual analysis to estimate how $X_{2i}$ might change if the ethnic identity of recipients were altered, thereby isolating the effect of ethnicity on organ quality allocation.

\subsection*{Outcome Fairness}
Outcome Fairness evaluates the long-term success of kidney transplants by examining graft lifespan across different ethnic groups. Graft lifespan refers to the duration from transplantation to graft failure, which occurs when the transplanted kidney ceases to function \citep{Wolfe2008}. In our model, let $Y_{1i}$ denote the graft lifespan (in days) for patient $i$. This criterion is critical for determining whether the allocation system results in fair long-term outcomes. 
It has been shown that graft lifespan is a key indicator of transplant success and patient quality of life \citep{Hariharan2021}. In Section~\ref{sec:3.3}, we conduct a competing risks analysis to model graft failure and rejection risks, allowing us to identify disparities in graft lifespan across different ethnic groups.

\subsection{Data Envelopment Analysis Framework}
\label{sec:2.2}

Our study employs Data Envelopment Analysis (DEA) to evaluate kidney allocation fairness across multiple criteria. This approach, originally developed by \citet{Charnes1978}, allows for a comprehensive analysis of efficiency in converting multiple inputs into outputs. In our context, we aim to simultaneously minimize certain criteria, such as waitlist duration and LKDPI score, while maximizing others, such as graft lifespan. This objective motivates our use of the hyperbolic graph efficiency measure in our DEA framework, an extension proposed by \citet{Fare1985}, where we treat the criteria to be minimized as inputs and those to be maximized as outputs.

Let $\{(X_i, Y_i)\}_{i=1}^n$ represent the set of input-output pairs for \(n\) patients, where \(X_i = (x_{1i}, \ldots, x_{pi}) \in \mathbb{R}_+^p\) denotes the \(p\) inputs and \(Y_i = (y_{1i}, \ldots, y_{qi}) \in \mathbb{R}_+^q\)  denotes the \(q\) outputs for patient \(i=1,\ldots,n\). In our kidney allocation analysis, we consider two key pre-transplant factors as inputs, waitlist duration and LKDPI score, represented by \(X_i = (x_{1i}, x_{2i})\) with $p=2$. The single output \(Y_i = y_{1i}\) with $q=1$ denotes graft lifespan. This structure enables the evaluation of how efficiently the kidney allocation system converts these pre-transplant factors into prolonged graft survival. 

Central to the DEA framework is the production possibility set
$$
\mathcal{T} = \left\{(X, Y) \in \mathbb{R}_+^p \times \mathbb{R}_+^q \ \middle| \ X \geq \sum_{j=1}^{n} \lambda_j X_j, \ Y \leq \sum_{j=1}^{n} \lambda_j Y_j, \ \sum_{j=1}^n \lambda_j = 1, \ \lambda_j \geq 0 \right\}.
$$

\noindent which represents the convex hull of all observed transplants. To account for and control structural differences in patient background characteristics, we extend the classical DEA framework using the conditional efficiency formulation of \citet{buadin2012measure}. Let $Z_i \in \mathbb{R}^r$ be a vector of exogenous covariates for $i$, such as education level, UNOS region, and citizenship status. To incorporate these covariates into the DEA framework, we define the covariate-conditional production set for each patient $i$ as
\begin{equation}
\mathcal{T}_{Z_i} = \left\{ 
(X, Y) \in \mathbb{R}_+^2 \times \mathbb{R}_+ \middle| 
X \geq \sum_{j \in \mathcal{I}_i} \lambda_j X_j,\ 
Y \leq \sum_{j \in \mathcal{I}_i} \lambda_j Y_j,\
\sum_{j \in \mathcal{I}_i} \lambda_j = 1,\ 
\lambda_j \geq 0
\right\},
\label{eq:conditional_production_set}
\end{equation}
where $K_h(Z_j, Z_i)$ is a smoothing kernel that measures similarity between the covariates of patients $j$ and $i$ and $\mathcal{I}_i = \bigl\{ j \in \{1,\dots,n\}: K_h(Z_j, Z_i) > 0 \bigr\}$. The kernel is constructed as
\begin{equation}
\scalebox{0.95}{$K_h(Z_j, Z_i) = \prod_{k \in \mathcal{C}} \left[ \mathbb{I}(Z_{jk} = Z_{ik}) + \left(1 - \mathbb{I}(Z_{jk} = Z_{ik})\right) e^{-1/h^2} \right] \cdot \exp\left( -\frac{\| Z_j^{\text{cont}} - Z_i^{\text{cont}} \|^2}{2h^2} \right),$}
\label{eq:kernel}
\end{equation}

\noindent where $\mathcal{C}$ is the index set of categorical variables and $Z^{\text{cont}}$ denotes the continuous subset of $Z$. The bandwidth parameter $h$ controls the degree of localization in the conditional efficiency estimation and is selected as $h = n^{-1/(r+4)}$, which represents an optimal bandwidth choice that achieves the best convergence rates for conditional DEA estimators \citep{buadin2012measure}. This construction softly localizes the frontier to account for environmental constraints, enabling a fairer comparison across patients with differing backgrounds.

The efficiency of a KPD transplant is quantified by the conditional hyperbolic efficiency score
\begin{equation}
 \theta_i(Z_i) = \min \left\{ \theta > 0 \,\middle|\, (\theta X_i, \theta^{-1} Y_i) \in \mathcal{T}_{Z_i} \right\},  
 \label{eq:cond_efficiency}
\end{equation}
which enables simultaneous and proportional adjustment of inputs and outputs needed to reach the frontier. The hyperbolic measure is particularly well-suited to the kidney allocation setting, as it allows for the simultaneous reduction of inputs (waitlist time, LKDPI) and increase of outputs (graft survival). This approach provides an intuitive interpretation as a single scalar value, facilitating comparisons across patient groups.

\color{black}
The efficiency frontier is a key component of our DEA framework, representing the set of input–output combinations $(X, Y) \in \mathcal{T}_{Z_i}$ for which no other feasible point satisfies $X' \leq X$ and $Y' \geq Y$ with at least one strict inequality. For any patient not on this frontier ($\theta_i(Z_i)<1$), the value of $\theta_i(Z_i)$ quantifies the proportional decrease in inputs and increase in output required for the observed transplant to reach the frontier.

\begin{figure}[tb]
\centering
\includegraphics[width=0.9\linewidth]{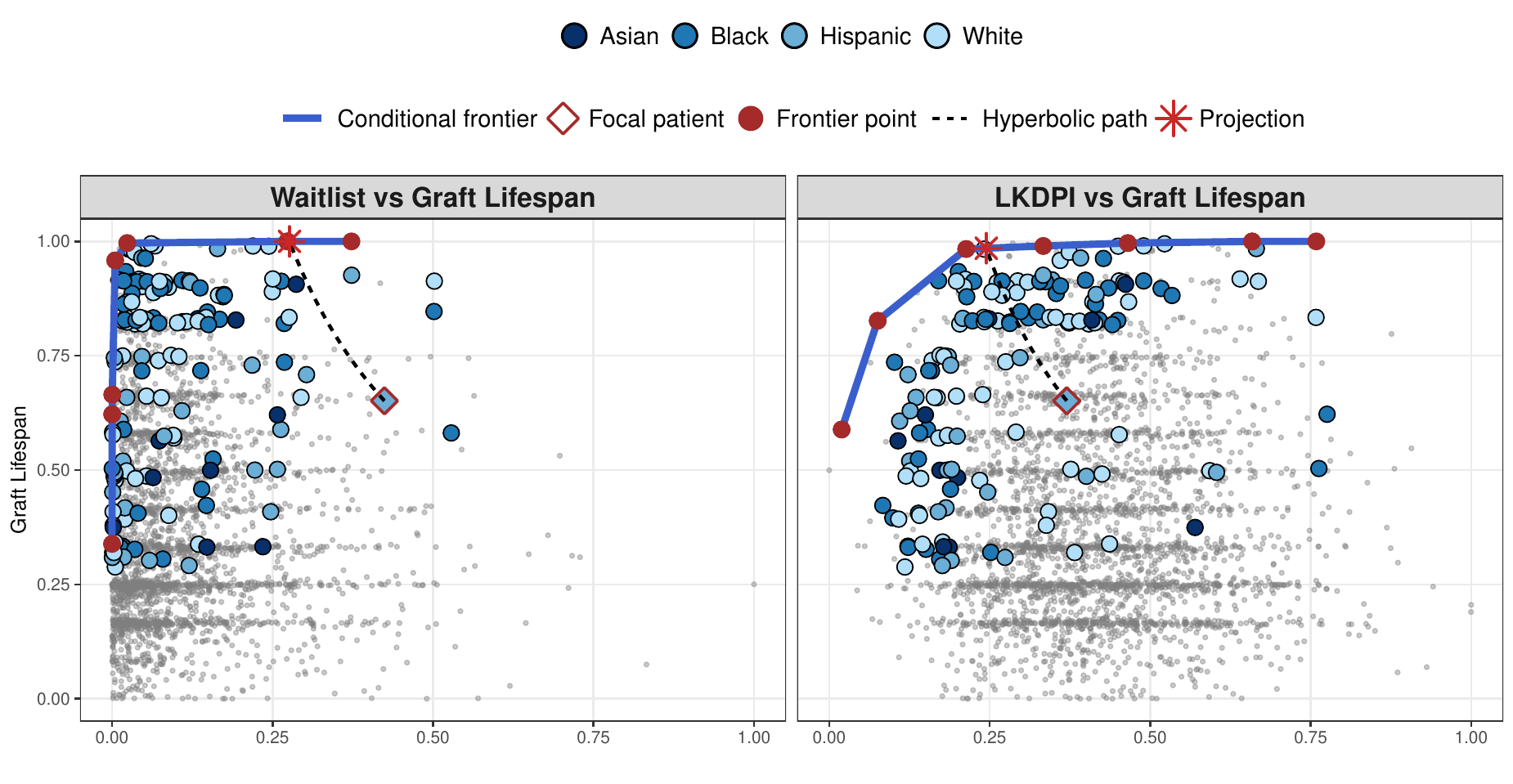}
\caption{\color{black}Illustration of the conditional DEA framework for a single focal patient. Each panel plots one input (waitlist duration or LKDPI) against the output (graft lifespan). Small points represent all observations in the sample, and the focal patient is marked with a diamond. Filled circular points indicate the focal patient's local reference set, consisting of patients with similar covariate profiles $Z$. The solid curve represents the covariate-conditional efficiency frontier estimated from this local reference set; points marked along this curve correspond to observations lying on the frontier. The dashed curve shows the hyperbolic trajectory along which the focal patient's input–output pair is scaled by $\theta$ until it intersects the frontier at the conditional efficiency target $\theta_i(Z_i)=1$.}
\label{fig:frontier}
\end{figure}

Figure~\ref{fig:frontier} illustrates how the conditional frontier and hyperbolic efficiency score are constructed for a single focal patient. To make the geometry visually interpretable, the inputs and output in these panels have been rescaled to the unit interval via min–max normalization, though this scaling does not affect the DEA estimates. Each panel displays all transplants as background points, highlighting one representative patient and the subset of reference units identified using covariate similarity. The solid curve depicts the patient-specific covariate-conditional frontier estimated from this local reference set. The dashed curve represents the hyperbolic path the patient's input–output combination would follow if proportionally scaled by $\theta$ until it reaches the conditional frontier at $\theta=1$, marked by a star. Although every patient has an individualized conditional frontier, we present a single example to maintain clarity and to demonstrate the geometric interpretation of $\theta_i(Z_i)$.
\color{black}

Incorporating the hyperbolic graph efficiency measure into the conditional DEA framework results in the following optimization problem for the efficiency of patient $i$:
\begin{equation}
\label{eq:dea_model}
\begin{aligned}
\min_{\theta, \lambda_j} \  \theta \quad\quad
\text{s.t.} \quad & \theta X_i \geq \sum_{j \in \mathcal{I}_i} \lambda_j X_j, \quad \frac{1}{\theta} Y_i \leq \sum_{j \in \mathcal{I}_i} \lambda_j Y_j, \\
& \sum_{j \in \mathcal{I}_i} \lambda_j = 1, \quad \lambda_j \geq 0,
\end{aligned}
\end{equation}
where $\mathcal{I}_i = \left\{ j \in \{1,\dots,n\} : K_h(Z_j, Z_i) > 0 \right\}$ denotes the locally weighted set of comparison units for patient $i$. These constraints ensure that each patient is only compared to peers with similar covariate profiles.

This framework enables us to compare the allocation efficiency across patients from different demographic groups on a common scale. By analyzing the distribution of efficiency scores across groups, we can identify disparities in the efficiency of allocation processes. For instance, consistently lower efficiency scores in one group may indicate less favorable transplantation outcomes relative to the efficient frontier, suggesting potential interventions to improve outcomes.

\subsection{Group-Conditional Uncertainty Quantification under Reference Frontier Mapping}
\label{sec:uq-rfm}

Uncertainty quantification provides confidence measures for the conditional efficiency scores computed in \eqref{eq:dea_model}, accounting for variability in input-output measurement and sampling. This is essential for evaluating fairness across ethnic groups by determining the statistical significance of efficiency score differences. To this end, we employ conformal prediction methods, which provide finite-sample coverage guarantees without relying on distributional assumptions and are well-suited for complex, non-linear relationships.

A central challenge in applying conformal prediction to DEA is that efficiency scores $\{\theta_i\}_{i=1}^n$ are computed relative to a frontier estimated from the full dataset. This induces statistical dependence between the construction of the frontier and the observations, violating the i.i.d. assumptions required for conformal inference. To resolve this, we propose a data-splitting framework, \textit{Reference Frontier Mapping} (RFM), which enables valid and equitable uncertainty quantification for conditional DEA efficiency scores. RFM separates the roles of frontier construction and score evaluation by partitioning the data into two disjoint subsets: a reference set $\mathcal{D}_\text{ref}$, used to estimate a covariate-conditional production frontier, and an evaluation set $\mathcal{D}_\text{eval}$, on which prediction intervals are computed.

The reference set is chosen to be informative about the boundary of the production possibility set, while the remaining observations form $\mathcal{D}_\text{eval}$. The targeted sampling and frontier-relevance scoring procedure used to construct $\mathcal{D}_\text{ref}$ is detailed in Supplementary Material Section~S1.1 \citep{supplement}. For each $i \in \mathcal{D}_\text{eval}$, we estimate a covariate-conditional production set $\mathcal{T}_{Z_i}$ using only reference units weighted by the kernel $K_h(Z_j,Z_i)$ in \eqref{eq:conditional_production_set}. This yields a locally adaptive frontier reflecting the structural and demographic context of KPD transplant patient $i$. The corresponding conditional efficiency score is then computed by solving \eqref{eq:dea_model}, which can be expressed as the functional mapping
$
\theta_i = \varphi_{\mathcal{T}_{Z_i}}(X_i, Y_i),
$
where $\varphi_{\mathcal{T}_{Z_i}} : \mathbb{R}^p \times \mathbb{R}^q \to \mathbb{R}$ denotes the DEA operator defined by the covariate-conditional frontier. Once $\mathcal{D}_{\text{ref}}$ is fixed, the evaluation triplets $(X_i,Y_i,\theta_i)$ are conditionally independent of frontier construction and may be treated as conditionally i.i.d. for conformal inference.

To obtain group-conditional uncertainty guarantees, we apply the method of \citet{Gibbs2023}, which ensures that prediction intervals achieve the target coverage level within each ethnic group. In brief, we estimate group-specific conditional means of $\theta_i$, compute conformity scores based on deviations from these estimates, and then invert a conformal calibration step to obtain prediction intervals satisfying
\begin{equation*}
\mathbb{P}\left(\theta_{n+1} \in \hat{C}(X_{n+1}, Y_{n+1}) \mid (X_{n+1}, Y_{n+1}) \in g\right) = 1 - \alpha,\quad \forall g \in \mathcal{G}.    
\end{equation*}
This is a group-specific coverage guarantee: conditional on belonging to group $g$, the interval contains the efficiency score with probability $1-\alpha$. The detailed optimization formulation, function-class specification, and randomized conformal prediction-set construction, together with the proof of the corresponding coverage theorem stated below, are provided in Sections~S1.2 and S2.1 of the Supplementary Material \citep{supplement}.

\begin{theorem}
\label{thm:conformal_coverage}
Let $\mathcal{G}$ denote a set of groups, and let ${(X_i, Y_i, Z_i)}_{i=1}^n$ be i.i.d. samples from a joint distribution $\mathbb{P}_{X,Y,Z}$, where $X_i \in \mathbb{R}^p$, $Y_i \in \mathbb{R}^q$, and $Z_i \in \mathbb{R}^r$. Let $\mathcal{D}_\text{ref}$ be a fixed reference set used to compute conditional efficiency scores via the mapping $\theta_i = \varphi_{\mathcal{T}_{Z_i}}(X_i, Y_i)$, where $\varphi_{\mathcal{T}_{Z_i}} : \mathbb{R}^p \times \mathbb{R}^q \to \mathbb{R}$ denotes the conditional DEA operator defined by $\mathcal{D}_\text{ref}$ and kernel smoothing. Then for any new sample $(X_{n+1}, Y_{n+1}, Z_{n+1}) \sim \mathbb{P}_{X,Y,Z}$, the randomized conformal prediction set $\hat{C}_\text{rand.}(X_{n+1}, Y_{n+1})$ satisfies
$$
\mathbb{P}\left(\theta_{n+1} \in \hat{C}_\text{rand.}(X_{n+1}, Y_{n+1}) \mid (X_{n+1}, Y_{n+1}) \in g\right) = 1 - \alpha, \quad \forall g \in \mathcal{G},
$$
\noindent where $\theta_{n+1} = \varphi_{\mathcal{T}_{Z_{n+1}}}(X_{n+1}, Y_{n+1})$ and $\alpha$ is the target miscoverage rate.
\end{theorem}

\noindent The efficiency score $\theta_{n+1}$ in Theorem~\ref{thm:conformal_coverage} represents the patient's conditional DEA efficiency relative to the fixed reference frontier $\mathcal{D}_\text{ref}$. This formulation does not require $(X_{n+1}, Y_{n+1})$ to correspond to future arrivals in a dynamic KPD system; instead, it provides uncertainty quantification for any new or held-out observation relative to a clearly defined historical benchmark. Each patient's efficiency score reflects performance relative to what was empirically achievable during their treatment period using completed transplants with similar covariate profiles. Although our analysis focuses on completed transplants within fixed time windows, the RFM framework can be extended to dynamic settings by periodically updating $\mathcal{D}_\text{ref}$ using rolling historical data. The resulting prediction set $\hat{C}(X_{n+1}, Y_{n+1})$  therefore provides rigorous, group-conditional uncertainty quantification for any patient-donor pair evaluated against the RFM frontier, enabling valid assessment of efficiency disparities across ethnic groups in both retrospective and prospective settings.

\section{Preliminary Analyses and Fairness Criteria}
\label{sec:3}
\subsection{Exploratory Data Analysis}
\label{sec:3.1}

The United Network for Organ Sharing (UNOS) operates the Organ Procurement and Transplantation Network (OPTN), which has collected organ transplant data since 1987 and serves as one of the most comprehensive national repositories for transplant data in the United States. Focusing specifically on kidney transplants conducted through KPD programs, we treat the UNOS registry as a representative sample of historical KPD transplants in the United States for our analysis. To provide a more accurate representation of the current landscape, we analyzed kidney transplant records from 2010 to 2019, focusing exclusively on the 6,365 patients who successfully received KPD transplants during this period, ensuring complete observation of all time-to-event outcomes. For each transplant, the UNOS data provides comprehensive information on a wide range of clinical and demographic variables, including recipient and donor age, ethnicity, blood type, panel reactive antibody (PRA) levels, and waitlist duration (measured as the time from initial waitlist registration to KPD transplant date). This enables analysis of both pre-transplant factors (e.g., waitlist duration) and post-transplant outcomes (e.g., graft survival time). 
Table \ref{tab:summary_stats} presents summary statistics for key variables across different ethnic groups.

\begin{table}[tbh]
\centering
\caption{Summary Statistics of Key Variables by Ethnic Group}
\label{tab:summary_stats}
\resizebox{0.99\columnwidth}{!}{
\begin{tabular}{lcccc}
\hline
Variable & Asian & Black & Hispanic & White \\
\hline
\textbf{Waitlist Duration (days)} &  &  &  &  \\
\quad Mean (SD) & 685.4 (611.3) & 691.5 (691.4) & 646.1 (698.5) & 514.3 (528.8) \\
\quad Median [IQR] & 465 [228, 1064] & 460 [203, 944] & 398.5 [140, 892] & 348 [166, 678] \\

\textbf{LKDPI} & & & & \\
\quad Mean (SD) & -8.30 (21.2) & -2.65 (20.8) & -7.67 (19.4) & -4.56 (20.6) \\
\quad Median [IQR] & -9.83 [-23.43, 4.11] & -4.37 [-17.69, 10.01] & -8.97 [-21.63, 5.09] & -6.45 [-19.05, 8.55] \\

\textbf{Graft Survival (days)} & & & & \\
\quad Mean (SD) & 1684.2 (881.2) & 1744.0 (990.3) & 1610.5 (895.5) & 1782.7 (978.6) \\
\quad Median [IQR] & 1459.0 [1090, 2197] & 1462.5 [1045.75, 2430.5] & 1418.5 [935.75,  2192.5] & 1486 [1071, 2513] \\
\hline
\end{tabular}}
\end{table}

These summary statistics show differences across ethnic groups in key transplantation metrics. Black patients, on average, experience the longest waitlist durations compared to other ethnic groups, with a mean of 691.5 days, followed closely by Asian (685.4 days) and Hispanic patients (646.1 days). White patients have the shortest mean waitlist duration of 514.3 days, substantially lower than other groups. Black patients receive kidneys with higher mean LKDPI scores (-2.65) in comparison to other groups. In terms of graft survival, White patients have the longest mean survival time (1782.7 days or approximately 4.9 years), followed closely by Black patients (1744.0 days or 4.8 years). Asian and Hispanic patients have shorter mean graft survival times (1684.2 and 1610.5 days or about 4.6 and 4.4 years, respectively). These observations suggest potential disparities in the kidney allocation process, with White patients appearing to have shorter waitlist duration and better graft survival, while other patients face longer wait times and shorter graft survival but have comparable LKDPI scores. These findings warrant further in-depth analysis to understand the underlying factors contributing to these differences across ethnic groups.

\label{sec:end_3.1}

\subsection{Data Preprocessing: Resampling and Relative Measures}
\label{sec:3.2}

Our data preparation proceeds in two stages: first, addressing group representation imbalance through stratified resampling; and second, defining fairness measures in relative terms rather than absolute terms.

\begin{figure}[tbh]
    \centering
    \includegraphics[width=0.7\textwidth]{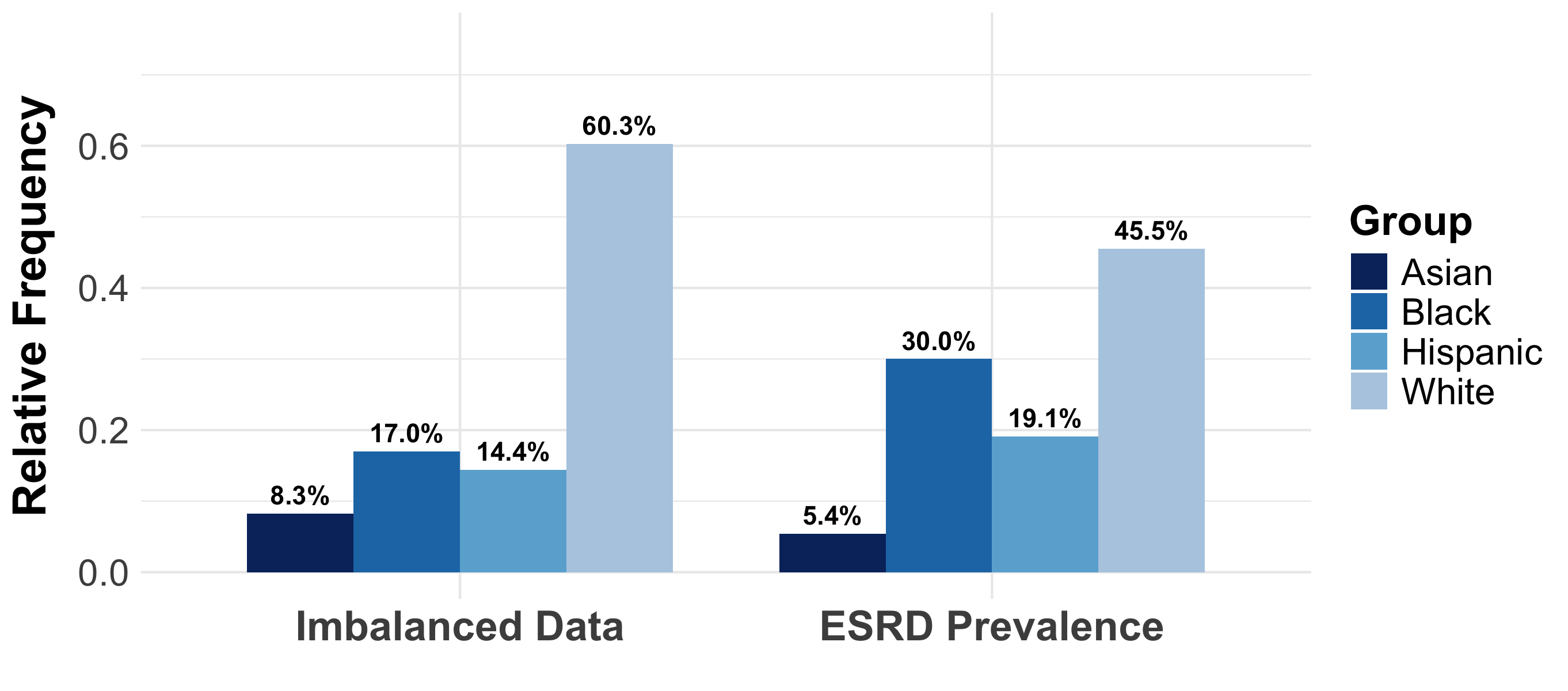}
    \caption{Comparison of the Distribution of Ethnic Groups in the 2019 Subset of the Imbalanced Data and the 2019 ESRD Prevalence in the U.S.}
    \label{fig:ethnic_distribution_2019}
\end{figure}

To illustrate the magnitude of group imbalance in our data, we compare the ethnic composition of patients who entered the KPD system in 2019 with the national ESRD prevalence for that year. According to the United States Renal Data System (USRDS) Annual Report \citep{USRDS2024}, the 2019 ESRD prevalence across major ethnic groups in the U.S. was: Asian (5.4\%), Black (30.0\%), Hispanic (19.1\%), and White (45.5\%). As shown in Figure~\ref{fig:ethnic_distribution_2019}, the empirical distribution of our 2019 KPD data deviates substantially from these population benchmarks. Similar imbalances are observed in other years as well, motivating the need for debiasing. Group representation disparities can compromise the validity of efficiency comparisons by introducing systematic bias into the DEA estimates. It has been shown that class imbalance can significantly affect the performance of learning algorithms, leading to biased predictions favoring the majority class \citep{He2009}. In our setting, this results in biased empirical efficiency estimates \(\hat{\theta}_g\) for group \(g\), relative to the population-level quantity of interest
$$
\theta_g^* := \mathbb{E}_{(X,Y,Z) \sim P_g}\left[\theta(X,Y \mid Z)\right],
$$
where $\theta(X, Y \mid Z)$ denotes the conditional DEA efficiency score of a patient with inputs \(X\), outputs \(Y\), and covariates \(Z\), evaluated relative to the covariate-specific production possibility set \(\mathcal{T}_Z \subset \mathbb{R}^{p+q}\). The distribution \(P_g\) denotes the joint distribution of \((X, Y, Z)\) given group membership \(\mathcal{G} = g\), and is supported on the union of the covariate-conditional production sets. That is,
$P_g(A) := \mathbb{P}((X,Y,Z) \in A \mid \mathcal{G} = g),$ for all measurable  $A \subseteq \mathbb{R}_+^p \times \mathbb{R}_+^q \times \mathbb{R}^r.$

To assess disparities in efficiency scores across ethnic groups in a manner that accurately reflects the demographic composition of patients affected by ESRD, we treat the U.S. ESRD patient population as the reference population to ensure fairness. Consequently, our resampling strategy aims to align the group proportions in the data with year-specific ESRD prevalence statistics from the USRDS by performing stratified resampling separately for each calendar year. For patients who entered the KPD system in year \(t\), we resample according to the population distribution of ESRD patients in that year. For each group \(g \in \mathcal{G}\) and year \(t\), let \(p_{g,t}\) be the ESRD prevalence proportion and \(n_{g,t}\) the number of observed patients in the original data. To avoid oversampling, we first compute the maximum feasible resampling size for year \(t\) as \(N_t = \min_{g \in \mathcal{G}} \left\lfloor n_{g,t}/p_{g,t} \right\rfloor\), where \( \lfloor \cdot \rfloor \) denotes the floor function. Then, within each group \(g\), we sample \(n_{g,t}' = \lfloor p_{g,t} \cdot N_t \rceil\) observations without replacement, where \(\lfloor \cdot \rceil \) denotes rounding to the nearest integer. This procedure yields a resampled dataset whose ethnic composition more closely reflects the year-specific ESRD population structure. Let \(n = \sum_{t} \sum_{g \in \mathcal{G}} n_{g,t}\) denote the total number of observations in the original dataset, and let \(n' = \sum_{t} \sum_{g \in \mathcal{G}} n_{g,t}' \leq n\) denote the total sample size after resampling. The resulting post-resampling group distribution is summarized in Table~\ref{tab:PostBalance}.
\begin{table}[tbh]
\centering
\caption{Distribution of Patients by Ethnicity — Post-Resampling}
\label{tab:PostBalance}
\begin{tabular}{lcc}
\hline
\textbf{Ethnicity} & \textbf{Number of Patients} & \textbf{Proportion of Total (\%)} \\
\hline
Asian    & 161  & 4.97\% \\
Black    & 1004  & 30.99\% \\
Hispanic & 582  & 17.96\% \\
White    & 1,493 & 46.08\% \\
\hline
Total    & 3,240 & \\
\hline
\end{tabular}
\end{table}

Let \(\hat{\theta}_g\) and \(\hat{\theta}_g'\) denote the empirical mean efficiency scores for group \(g\) computed from the original and resampled datasets, respectively, obtained by the DEA estimation problem in \eqref{eq:dea_model} and averaging the resulting scores across individuals in group \(g\). Theorem~\ref{thm:consistency-bias} below formalizes the bias reduction achieved through the resampling procedure.

\begin{theorem}
\label{thm:consistency-bias}
Let $\mathcal{T}_Z \subset \mathbb{R}^{p+q}$ denote the covariate-conditional production set defined by the DEA kernel-based estimator for each patient with covariate profile $Z \in \mathbb{R}^r$. Assume that $\mathcal{T}_Z$ is compact and possesses a $C^2$-smooth frontier for all $Z$, and that the smoothing kernel in \eqref{eq:kernel} is Lipschitz continuous in $Z$. For each group \( g \in \mathcal{G} \), let \( \theta_g^*\) denote the true conditional efficiency, \( \hat{\theta}_g \) the empirical mean efficiency score from the imbalanced dataset of size \( n \), and \( \hat{\theta}_g' \) the corresponding estimate from the resampled dataset of size $n'$, where \( n' \leq n \). Let $\delta > 0$ be a constant representing the desired precision in estimating the true efficiency scores, and define
\(
K \;:=\; \sup_{g\in\mathcal{G}}
\Bigl\{
\limsup_{n\to\infty}\,
n^{4/((r+4)(p+q))} 
\,\mathbb{E}\bigl[\bigl|\hat{\theta}_g - \theta_g^*\bigr|\bigr]
\Bigr\}
\)
as a constant characterizing the convergence rate of the estimators. Then, for $ n' \;\ge\; \max\Bigl\{\,\bigl(\frac{K}{\delta}\bigr)^{\frac{(r+4)(p+q)}{4}}, \;\min_{g\in\mathcal{G}}\tfrac{1}{p_g}\Bigr\} $, we have
$$
|\text{Bias}(\hat{\theta}_{g}')| \leq |\text{Bias}(\hat{\theta}_{g})| \quad \forall g \in \mathcal{G},
$$
where $\text{Bias}(\hat{\theta}_{g}) := \mathbb{E}[\hat{\theta}_{g}] - \theta^*_g$.
\end{theorem}
\label{sec:end_resampling}
\noindent The constant \(K\) in Theorem~\ref{thm:consistency-bias} depends on the dimensionality \((p, q, r)\) of the input-output-covariate space and reflects both the curvature of the conditional frontier \(\partial \mathcal{T}_Z\) and the smoothness of the covariate kernel used to define the local neighborhood around each observation. Under the assumptions that the conditional production set \(\mathcal{T}_Z\) is compact with a twice continuously differentiable boundary and that the kernel function \(K_h(Z_j, Z)\) is Lipschitz continuous, conditional DEA estimators are known to be consistent with convergence rates of order \(O(n^{-4/((r+4)(p+q))})\), which accounts for the effective sample size \(nh^r\) \citep{buadin2012measure}. The true group-level efficiency score \(\theta_g^*\) is defined with respect to the U.S. ESRD patient population; thus, the empirical data used in our analysis must reflect the demographic composition of this population to provide unbiased and representative estimates.

\label{sec:score_centering}
Following the resampling step, we define the fairness criteria used in our analysis. For each KPD transplant $i$ performed in year $t$, we denote $X_{1i}^{(t)}$ as the observed waitlist duration, $X_{2i}^{(t)}$ as the observed LKDPI score, and $Y_{1i}^{(t)}$ as the observed graft lifespan. 

Our analysis employs two distinct but complementary approaches that require different data transformations. For the individual fairness analyses described in Section~\ref{sec:3.3}, we center each criterion around its year-specific sample mean to evaluate fairness relative to expected outcomes and account for temporal changes in the transplant landscape. Specifically, we define the centered fairness measures as
\[
\tilde{X}_{1i}^{(t)} := X_{1i}^{(t)} - \bar{X}_1^{(t)}, \quad
\tilde{X}_{2i}^{(t)} := X_{2i}^{(t)} - \bar{X}_2^{(t)}, \quad
\tilde{Y}_{1i}^{(t)} := Y_{1i}^{(t)} - \bar{Y}_1^{(t)},
\]
where $\bar{X}_1^{(t)}$, $\bar{X}_2^{(t)}$, and $\bar{Y}_1^{(t)}$ denote the sample means of waitlist duration, LKDPI, and graft lifespan, respectively, computed over all patients transplanted in year $t$ within the resampled dataset. This year-specific centering procedure ensures that fairness comparisons account for temporal changes in transplant practices, allocation policies, and patient populations. The centered fairness measures $\tilde{X}_{1i}^{(t)}$, $\tilde{X}_{2i}^{(t)}$, and $\tilde{Y}_{1i}^{(t)}$ quantify whether each KPD transplant achieved better or worse outcomes relative to the yearly sample mean, thereby enabling a relative assessment of priority fairness, access fairness, and outcome fairness in the individual analyses presented in Section~\ref{sec:3.3}. Throughout the analyses in Section~\ref{sec:3.3}, we use $X_1$, $X_2$, and $Y_1$ to refer to these year-centered fairness measures. Table~\ref{tab:centered_stats} presents summary statistics for these centered fairness criteria across ethnic groups, providing an overview of the magnitude and direction of group-specific deviations from the year-specific sample means.

\begin{table}[tbh]
\centering
\caption{Summary Statistics of Centered Fairness Measures by Ethnic Group}
\label{tab:centered_stats}
\resizebox{0.99\columnwidth}{!}{
\begin{tabular}{lcccc}
\hline
Variable & Asian & Black & Hispanic & White \\
\hline
\textbf{Priority Fairness} & & & & \\
\quad Mean (SD) & 83.6 (612.8) & 90.2 (690.3) & 44.5 (697.4) & -87.0 (528.4) \\
\quad Median [IQR] & -125.79 [-386.94, 448.17] & -142.69 [-392.80, 334.00] & -212.67 [-462.61, 294.13] & -251.35 [-434.83, 71.60] \\

\textbf{Access Fairness} & & & & \\
\quad Mean (SD) & -3.6 (21.1) & 2.1 (20.8) & -3.0 (19.4) & 0.2 (20.6) \\
\quad Median [IQR] & -4.48 [-18.84, 9.81] & 0.50 [-12.94, 14.57] & -4.50 [-16.76, 9.53] & -1.55 [-14.57, 13.40] \\

\textbf{Outcome Fairness} & & & & \\
\quad Mean (SD) & -0.8 (692.1) & -5.5 (712.8) & -94.0 (648.4) & 40.4 (684.4) \\
\quad Median [IQR] & 150.78 [-108.53, 332.06] & 99.06 [-167.64, 368.42] & -8.09 [-293.96, 291.70] & 158.92 [-129.53, 427.56] \\
\hline
\end{tabular}}
\end{table}

For the DEA framework described in Section~\ref{sec:2.2}, we preserve the natural input-output relationships essential for meaningful efficiency measurement by using the original observed values with minimal transformation. Specifically, we use the raw waitlist duration $X_{1i}^{(t)}$ and graft lifespan $Y_{1i}^{(t)}$ directly, while applying a constant shift to LKDPI scores equal to the absolute value of the minimum observed LKDPI value to ensure non-negativity as required by DEA. We then apply year-specific centering to the resulting DEA efficiency scores themselves, as described in Section~\ref{sec:4.1}, to account for temporal changes in the efficiency frontier while preserving the interpretability of relative performance measures.

\color{black}

\label{sec:end_3.2}

\subsection{Fairness Criteria Analysis}
\label{sec:3.3}
To thoroughly evaluate fairness in kidney allocation, we study three criteria: Priority Fairness (waitlist duration), Access Fairness (LKDPI score), and Outcome Fairness (graft lifespan). These detailed analyses provide a foundation for our subsequent Data Envelopment Analysis.

\subsubsection*{Priority Fairness: Mediation Analysis}

Priority fairness aims to identify and quantify disparities in waitlist durations across ethnic groups that cannot be explained solely by medical factors. To achieve this, we use mediation analysis to decompose the total effect of ethnicity into direct and indirect effect pathways \citep{MacKinnon2008}. This approach enables us to isolate disparities that persist even after accounting for observable patient characteristics.

To identify which variables to include as mediators or covariates in the mediation model, we implemented a structured two-stage hypothesis testing procedure. In the first stage, we assessed the association between each candidate variable $M_j \in M$ and the outcome $X_1$ (waitlist duration) using likelihood ratio tests within a generalized linear modeling framework. Each test compared a full model including $M_j$ to a reduced model excluding it, thereby evaluating the incremental explanatory power of $M_j$ conditional on the other variables. In the second stage, we tested whether each candidate variable $M_j$ was statistically associated with the treatment variable $G$ (ethnicity). Here, the choice of model depended on the structure of the mediator. We used linear models for continuous mediators, logistic regression for binary mediators, and multinomial logistic regression for categorical mediators with more than two levels. In each case, we compared a full model that included ethnicity as a predictor to a null model that excluded it, and computed likelihood ratio statistics to assess significance. To control for false discoveries across multiple comparisons, we applied the Benjamini–Hochberg procedure \citep{benjamini1995controlling} in both stages to adjust the resulting $p$-values. Candidate variables with BH-adjusted $p$-values less than 0.05 in both stages were classified as mediators, as they showed evidence of being influenced by ethnicity and of affecting waitlist duration. Variables significantly associated only with the outcome were classified as covariates. Variables not significantly associated with waitlist duration were excluded from the mediation model, as they did not contribute to explaining the variation in transplant priority.

Based on this classification procedure, we include the following variables as mediators in the final model: Recipient Age, ABO Blood Type, UNOS Transplant Region, Recipient Education, Recipient Citizenship Status, Recipient Employment Status, and Recipient Dialysis Status. These variables were significantly associated with both ethnicity and waitlist duration, indicating that they may lie along a causal pathway connecting the two. Additionally, we include the PRA Score as a covariate, as it is significantly associated with waitlist duration but not with ethnicity. All other candidate variables were excluded because they did not exhibit a significant association in the initial testing stage. Detailed results from both stages of the testing procedure, including test statistics and adjusted $p$-values for each candidate variable, are provided in Section S3.1 of the Supplementary Material \citep{supplement}.

We formulate the mediation model as a system of regression equations to estimate both the direct effects of ethnicity and the mediated (indirect) pathways through the selected variables. Let $G$ denote the categorical variable for ethnicity, and define $X_1$ as the observed waitlist duration. The set of mediators is denoted $M = (M_1, \dots, M_7)$, where each $M_m$ corresponds to one of the seven mediator variables listed above. Let $C$ denote the PRA covariate. The total, direct, and indirect effects are estimated using the following linear model.
\begin{equation*}
X_1 = \beta_0 + \sum_{g \in \mathcal{G} \setminus \{\text{White}\}} \beta_g \mathbb{I}(G = g) + \tau^\top M + \zeta C + \epsilon,
\end{equation*}
where $\beta_g$ represents the direct effect of ethnicity group $g$ on waitlist duration, $\tau$ is the vector of coefficients for the mediators $M$, and $\zeta$ is the coefficient for the covariate $C$.

Each mediator $M_m$ is regressed individually on the group indicator $G$ using the appropriate regression model, depending on its type (logistic for binary, multinomial for categorical, and linear for continuous mediators). For each mediator $M_m$, the fitted model takes the general form:
\begin{equation*}
M_m = \gamma_0 + \sum_{g \in \mathcal{G} \setminus \{\text{White}\}} \gamma_{g,m} \mathbb{I}(G = g) + \nu_m,
\end{equation*}
where $\gamma_{g,m}$ captures the effect of ethnicity $g$ on mediator $m$, and $\nu_m$ is the residual error term capturing variation in $M_m$ not explained by ethnicity. 

The indirect effect for each ethnic group $g$ is expressed as
\begin{equation*}
\text{IE}_g = \sum_{m} \tau_m \left( \mathbb{E}[M_m \mid G = g] - \mathbb{E}[M_m \mid G = \text{White}] \right),
\end{equation*}
and the direct effect is then defined as
\begin{equation*}
\hat{\beta}_g = \text{TE}_g - \text{IE}_g,
\end{equation*}
where $\text{TE}_g = \mathbb{E}[X_1 \mid G = g] - \mathbb{E}[X_1 \mid G = \text{White}]$ is the total effect. This formulation allows us to isolate disparities in waitlist duration attributable to ethnicity that remain after adjusting for all mediators and covariates included in the model. A significant $\hat{\beta}_g$ suggests that, even after accounting for the factors in our model, there remain substantial differences in waitlist duration for ethnic group $g$ compared to the White patient reference group. 

We estimate these effects using the \texttt{mma} package in \texttt{R} with 1000 bootstrap samples to obtain standard errors and confidence intervals. Results from this analysis are presented in Table~\ref{tab:wl_results_effects}, showing significant disparities in waitlist durations across ethnic groups. Substantial direct effects for all minority groups indicate unfairness not fully explained by the chosen mediators. The varying proportions of mediated effects suggest non-uniform mechanisms underlying these disparities. 

\begin{table}[tbh]
\centering
\caption[Summary of Mediation Model Effects]{Summary of direct, indirect, and total effects for each ethnic group. *** indicates $p$-value $< 0.001$.}
\label{tab:wl_results_effects}

\begin{tabular}{llccc}
\hline
\textbf{Patient Ethnicity} & \textbf{Effect Type} & \textbf{Estimate} & \textbf{SE} & \textbf{95\% CI} \\
\hline
\textit{Asian} & & & & \\
 & Direct Effect  ** & 132.433 & 50.391 &  [37.006, 228.527] \\
 & Indirect Effect * & 52.847 & 24.671 & [7.679, 104.372] \\
 & Total Effect *** & 185.280 & 49.967 &  [96.275, 285.998] \\
& {Proportion Mediated}  &  0.285 &  &   \\
\hline
\textit{Black} & & & & \\
 & Direct Effect *** & 187.601 & 26.442 &  [135.300, 240.471] \\
 & Indirect Effect  & -10.157 & 10.039 & [-29.608, 9.627] \\
 & Total Effect *** &  177.445 &  26.102 &  [128.225, 226.646] \\
& {Proportion Mediated}  &  0.057 &  &   \\
\hline
\textit{Hispanic} & & & & \\
 & Direct Effect *** & 151.760 & 34.889 &  [88.457, 221.877] \\
 & Indirect Effect  & -23.345 & 19.396 & [-61.872, 12.478] \\
 & Total Effect *** &  128.415 &  31.323 &  [66.728, 192.739] \\
& {Proportion Mediated}  &  0.182 &  &   \\
\hline
\end{tabular}
\end{table}

\color{black}

\subsubsection*{Access Fairness: Counterfactual Analysis with Random Forest}
\label{access_analysis_3.3}

Access fairness investigates potential inequalities in the quality of kidneys allocated across different ethnic groups, using the LKDPI score as a key metric. We employed a counterfactual analysis using a random forest algorithm to isolate the effect of ethnicity on LKDPI scores, while accounting for other influential factors. To implement our methodology with five-fold cross-validation, consider the framework,
\[\hat{X}_{2} = \frac{1}{N} \sum_{l=1}^{N} T_l (\mathcal{Z} ; \Theta_l),\]
\noindent where $\hat{X}_{2}$ denotes the predicted LKDPI score, $N = 500$ is the number of trees, $T_l$ is the $l^{th}$ tree's prediction, and $\mathcal{Z}$ represents the feature set encompassing donor clinical and demographic characteristics involved in the computation of LKDPI scores (age, eGFR, BMI, SBP, ethnicity, smoking history, hypertension, diabetes, hepatitis C status, HLA mismatches, donor-recipient gender pair, ABO incompatibility, and donor-recipient weight ratio), along with recipient socioeconomic covariates such as education, insurance type, employment status, and UNOS transplant region. The random parameters $\Theta_l$ for each tree $l$ govern the tree construction process by determining the subset of features randomly selected for evaluating splits at each node, with cross-validation determining that 12 features at each split yielded optimal model performance. Each tree is constructed by recursively splitting nodes, maximizing information gain at each split. Trees are grown to their maximum depth, with splitting continuing until each terminal node contains at least 5 samples, ensuring a balance between model complexity and predictive power.

To understand the factors influencing LKDPI scores, we conducted a variable importance analysis using permutation importance based on out-of-bag estimates \citep{Breiman2001}. This analysis shows that donor attributes in $\mathcal{Z}$ are the most significant factors influencing LKDPI scores. Specifically, donor age emerged as the most important feature, followed by donor sex, eGFR, Black ethnicity, smoking history, and systolic blood pressure. Other relevant features include donor-recipient gender match, HLA mismatches, ABO incompatibility, and the donor-to-recipient weight ratio. Among recipient-related and socioeconomic features, only marginal effects were observed, with variables such as education, insurance type, and citizenship status contributing little to overall predictive accuracy. Our model demonstrated strong performance with high predictive accuracy, yielding a coefficient of determination $R^2 =  0.985$. The model's performance is further quantified by the mean absolute error, $\text{MAE} = n^{-1}\sum_{i=1}^n |\hat{X}_{2,i} - X_{2,i}| = 1.793$, and the mean squared error, $\text{MSE} = n^{-1}\sum_{i=1}^n (\hat{X}_{2,i} - X_{2,i})^2 = 6.371$, where $\hat{X}_{2,i}$ is the predicted LKDPI score and $X_{2,i}$ is the actual LKDPI score for the $i$th observation. To isolate the effect of patient ethnicity, we created a counterfactual scenario in which the ethnicity of all recipients was set to `White', enabling the estimation of the effect of ethnicity on LKDPI scores.

\begin{table}[tbh]
\centering
\caption{Summary of LKDPI Score Disparities Across Ethnicities}
\label{tab:lkdpi_scores}
\begin{tabular}{lccc}
\hline
\textbf{Ethnicity} & \textbf{Mean LKDPI (Actual)} & \textbf{Mean LKDPI (Counterfactual)} & \textbf{Mean Diff.} \\
\hline
White & -0.0041 & - & -  \\
Asian & -2.974 & -2.8871 & 0.087  \\
Black & 1.782 & 1.697 & -0.0849 \\
Hispanic & -3.0295 & -2.947 & 0.0823  \\
\hline
\end{tabular}
\end{table}

The results of this analysis are summarized in Table \ref{tab:lkdpi_scores}, which shows minimal disparities in LKDPI scores across ethnic groups, with the largest difference observed in the Asian group (a mean difference of 0.087 between actual and counterfactual scenarios). While these differences are statistically significant (Kruskal-Wallis test, $p$-value < 0.001), their practical significance is negligible. Our variable importance analysis indicates that recipient ethnicity does not directly influence LKDPI scores; instead, donor characteristics are the primary factors. This suggests that disparities could arise if certain ethnic groups consistently receive kidneys from donors with specific attributes.

\label{access_analysis_3.3_end}

\subsubsection*{Outcome Fairness: Competing Risks Analysis}

Outcome fairness focuses on identifying disparities in graft lifespan. To analyze the effect of ethnicity on graft rejection while accounting for other causes of graft failure, we employ the Competing Risks Framework \citep{Fine1999}, which offers a comprehensive understanding of the risks associated with graft failure. 

In our framework, the subdistribution hazard function for cause $k$ is expressed by
\[ h_k(y_1 \mid G, W) = h_{0k}(y_1) \exp(\delta'G  + \gamma' W), \]
\noindent where $h_k(y_1 \mid G, W)$ is the subdistribution hazard for cause $k$ at time $y_1$ since transplantation, $h_{0k}(y_1)$ is the baseline subdistribution hazard for cause $k$, $G$ is a vector of effect-coded variables for ethnic groups (Asian, Black, Hispanic) using sum-to-zero contrasts, and $W$ is a vector of observed covariates including employment status at time of transplant and whether the patient received post-transplant rejection treatment. The coefficient vector $\delta$ captures the estimated deviations of each ethnic group from the overall sample mean subdistribution hazard for graft rejection, the primary event of interest, while treating all other causes of graft failure as competing risks and adjusting for clinical and socioeconomic confounders. The cause-specific cumulative incidence function for graft rejection is
\[ F_1(y_1) = \int_0^{y_1} h_1(u)\exp\left(-\int_0^u \bigl(h_1(s) + h_2(s)\bigr)\,ds\right)\,du, \]
where $Y_1$ represents the time to graft failure, $D$ is the event type (1 for rejection, 2 for other causes), and $h_1$ and $h_2$ are the cause-specific hazard functions for rejection and other causes, respectively.

\begin{table}[tbh]
\centering
\caption{Subdistribution Hazard Ratios for Graft Rejection Relative to Overall Sample Mean, Adjusted for Covariates $W$. Coefficients use sum-to-zero contrasts; the White coefficient is the negative sum of the other groups.}
\label{table:ethnic_disparities_graft_rejection1}
\begin{tabular}{lcccccc}
\hline
\textbf{Ethnicity} & \textbf{$\delta$} & \textbf{exp($\boldsymbol{\delta}$)} & \textbf{SE($\boldsymbol{\delta}$)} & \textbf{z} & \textbf{$p$-value} & \textbf{95\% CI for exp($\boldsymbol{\delta}$)} \\
\hline
Asian & -0.046 & 0.955 & 0.220 & -0.208 & 0.835 & [0.621, 1.470] \\
Black & -0.888 & 0.412 & 0.536 & -1.657 & 0.098 & [0.144, 1.177] \\
Hispanic & 0.727 & 2.068 & 0.212 & 3.428 & 0.001 & [1.365, 3.133] \\
White & 0.207 & 1.230 & 0.252 & 0.821 & 0.412 & [0.750, 2.016] \\
\hline
\end{tabular}
\end{table}

\begin{figure}[tbh]
\centering
\includegraphics[width=0.9\textwidth]{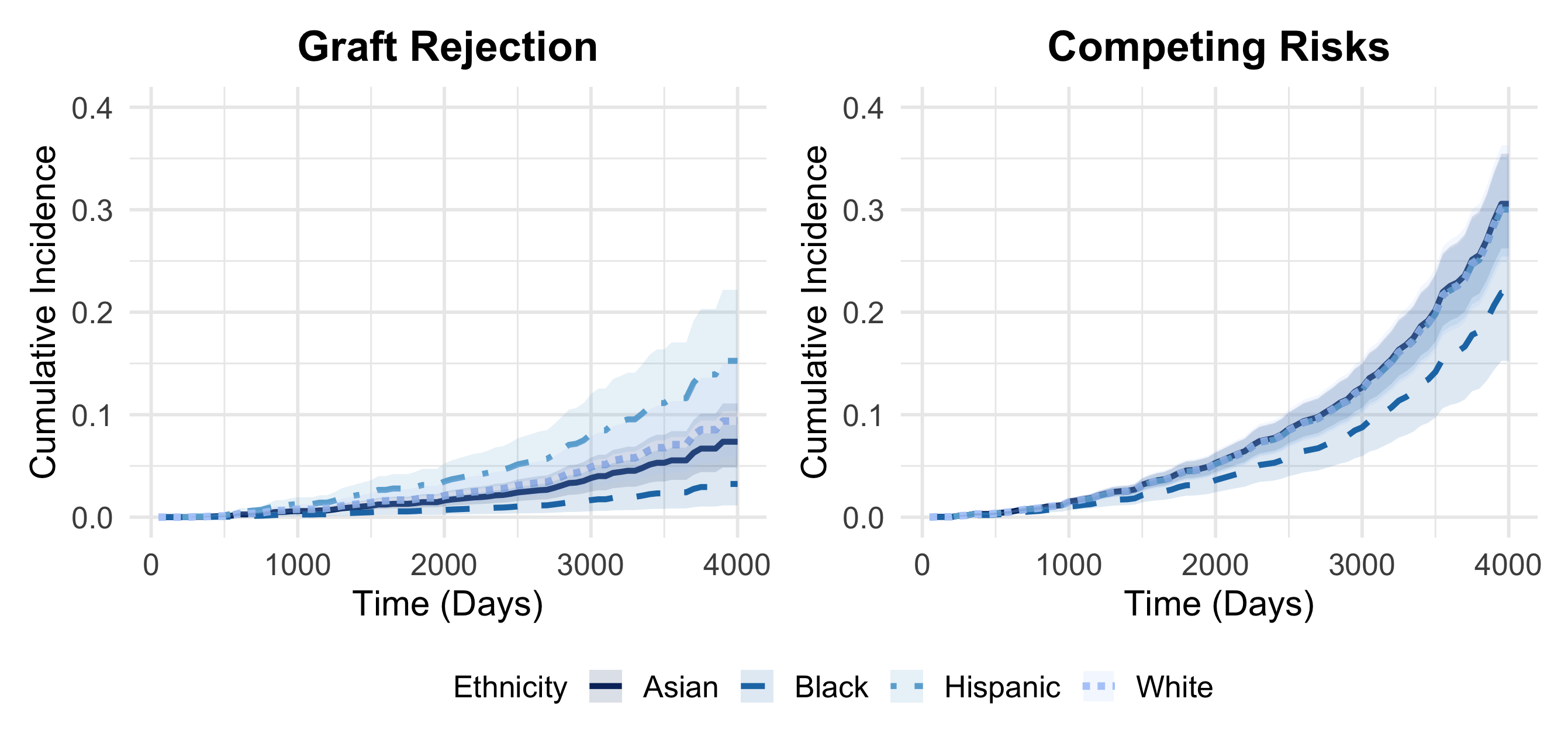}
\caption{Cumulative Incidence Functions for Graft Rejection and Competing Risks by Ethnicity}
\label{fig:cif_plot}
\end{figure}

Figure~\ref{fig:cif_plot} illustrates the cumulative incidence functions for graft rejection and competing risks, stratified by ethnicity and adjusted for observed covariates. Hispanic recipients exhibit the highest adjusted cumulative incidence of graft rejection throughout the post-transplant period, followed by White, Asian, and Black recipients. These differences are consistent with the subdistribution hazard ratios reported in Table~\ref{table:ethnic_disparities_graft_rejection1}, which indicate that Hispanic patients have a statistically significant elevation in rejection risk relative to the overall sample mean (HR = 2.07, 95\% CI: 1.37-3.13, p < 0.001). In contrast, Black recipients demonstrate the lowest risk of graft rejection relative to the sample average, though this difference does not reach statistical significance (HR = 0.41, 95\% CI: 0.14-1.18, p = 0.098). Asian and White recipients show risk profiles closer to the sample average, with neither group demonstrating statistically significant deviations from the population mean. In contrast, for graft failure attributable to causes other than rejection, the risk patterns differ considerably from those observed for immunologic failure. This divergence in risk profiles across failure types emphasizes the multifaceted nature of transplant outcomes. The findings suggest that Hispanic recipients are disproportionately affected by graft rejection even after adjustment for socioeconomic and post-transplant clinical factors, while Black recipients exhibit a lower risk of rejection-related graft loss compared to the study population average. Asian and White recipients demonstrate risk profiles that approximate the population mean for rejection-related graft failure.

\section{DEA Estimation and Inference}
\label{sec:4}
Given the complex and sometimes conflicting results from our individual fairness criteria analyses, we use DEA to gain deeper insights into fairness in kidney allocation. DEA enables the simultaneous consideration of multiple inputs and outputs, helping to identify overall disparities while accounting for the multifaceted nature of fairness in the allocation process.

\subsection{DEA Results}
\label{sec:4.1}
In applying the conditional DEA model defined in \eqref{eq:dea_model}, we incorporate waitlist duration and LKDPI score as inputs ($p=2$, $X_i \in \mathbb{R}_+^2$) and graft lifespan as the output ($q=1$, $Y_i \in \mathbb{R}_+$) for each patient $i$. The production frontier for each patient is localized using the covariate-conditional set $\mathcal{T}_{Z_i}$ as described in Section~\ref{sec:2.2}, enabling the resulting efficiency score $\theta_i(Z_i)$ to reflect performance relative to patients with similar socioeconomic and clinical backgrounds.

To assess disparities over time and adjust for temporal changes in the transplant landscape, we compute relative efficiency scores by centering each patient’s score against the mean efficiency of transplants performed in the same calendar year. Specifically, for each patient $i$ transplanted in year $t$, we define the centered efficiency score as $\tilde{\theta}_i := \theta_i - \bar{\theta}_t,$
where $\bar{\theta}_t$ is the mean conditional DEA score among all KPD recipients in year $t$. This centering procedure accounts for year-to-year variation and enables comparison of relative efficiency levels across groups.

\begin{table}[tbh]
\centering
\caption[Efficiency Scores and Prediction Intervals by Ethnic Group]{\raggedright Relative conditional DEA scores and group-conditional prediction intervals by ethnic group across cumulative time periods. Efficiency scores were estimated using the \texttt{Benchmarking} package in \texttt{R}.}
\label{tab:intervals}

\begin{tabular}{llccc}
\hline
Period & Group & Mean Relative Efficiency & Prediction Interval ($95\%$) & Coverage\\
\hline
\multirow{4}{*}{2010-2013} 
 & Asian & $-0.053$ & [$-0.6581,\ 0.3767$] & $0.956$ \\
 & Black & $-0.0024$ & [$-0.5655,\ 0.3609$] & $0.953$ \\
 & Hispanic & $-0.0113$ & [$-0.5709,\ 0.3827$] & $0.954$ \\
 & White & $0.0074$ & [$-0.5593,\ 0.3754$] & $0.952$ \\
\hline
\multirow{4}{*}{2010-2016} 
 & Asian & $-0.0073$ & [$-0.5300,\ 0.3773$] & $0.952$ \\
 & Black & $-0.0060$ & [$-0.5136,\ 0.3459$] & $0.949$ \\
 & Hispanic & $-0.0152$ & [$-0.4877,\ 0.3364$] & $0.950$ \\
 & White & $0.0104$ & [$-0.4911,\ 0.3236$] & $0.950$ \\
\hline
\multirow{4}{*}{2010-2019} 
 & Asian & $0.0055$ & [$-0.4879,\ 0.3551$] & $0.956$ \\
 & Black & $-0.0071$ & [$-0.4874,\ 0.3363$] & $0.950$ \\
 & Hispanic & $-0.0047$ & [$-0.4377, 0.3279$] & $0.950$ \\
 & White & $0.0066$ & [$-0.4466,\ 0.3189$] & $0.950$ \\
\hline
\end{tabular}
\end{table}

\begin{figure}[tbh]
    \centering
    \includegraphics[width=0.9\textwidth]{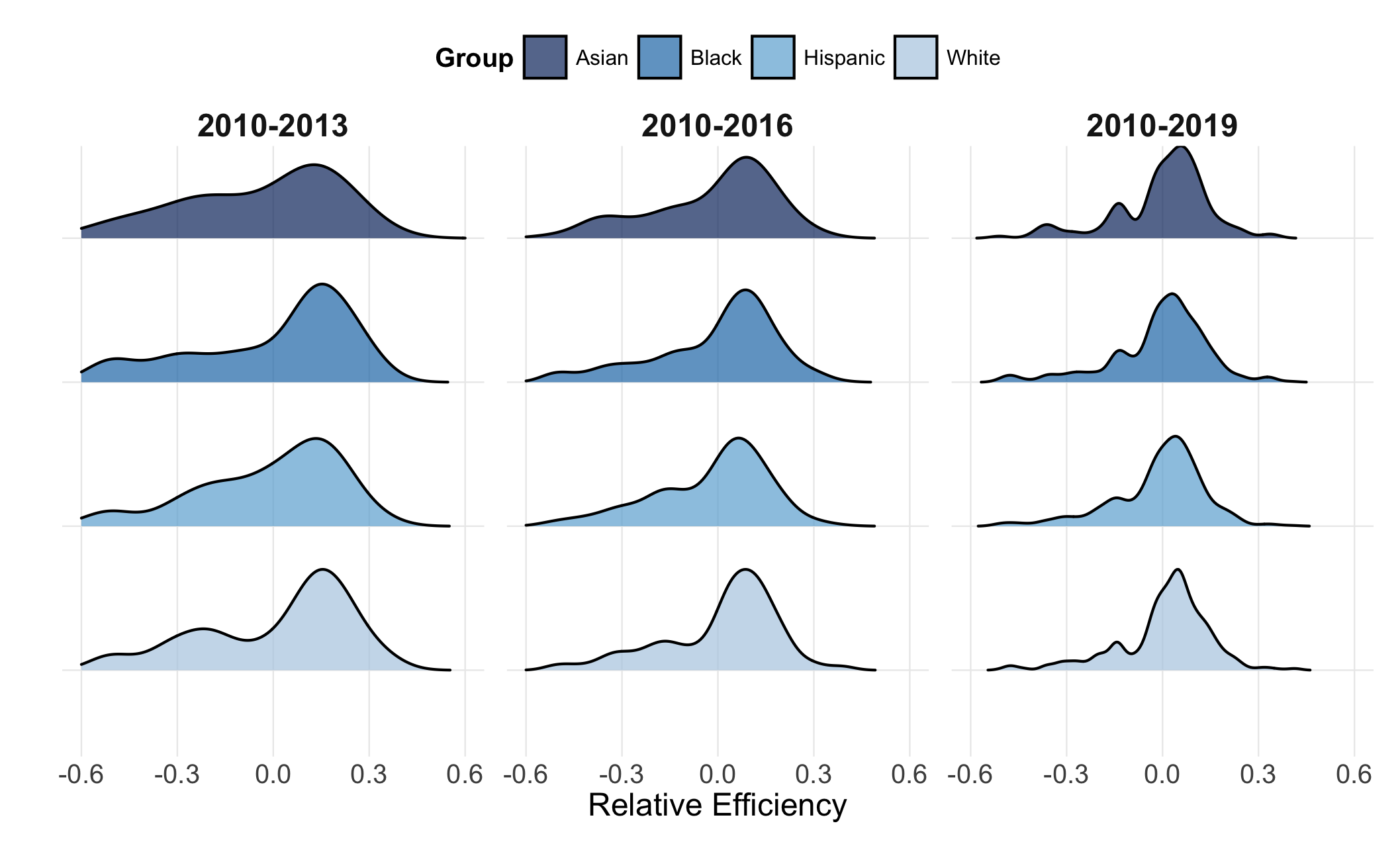}
    \caption{Distribution of Relative Efficiency Scores by Ethnic Group.}
    \label{fig:ridge}
\end{figure}

The results in Table~\ref{tab:intervals}, together with the distributions shown in Figure~\ref{fig:ridge}, provide a comprehensive view of efficiency across three cumulative time periods (2010–2013, 2010–2016, and 2010–2019). For each ethnic group and period, we report the mean centered efficiency score $\tilde{\theta}_i$, along with $95\%$ prediction intervals and empirical coverage rates, all computed using group-conditional conformal prediction under the Reference Frontier Mapping (RFM) procedure introduced in Section~\ref{sec:uq-rfm} and averaged over 100 random train–calibration–test splits to account for sampling variability. These prediction intervals ensure that, for each group, the interval contains the true efficiency score with $95\%$ probability, providing robust and equitable uncertainty quantification across groups. The mean relative efficiency scores are generally close to zero across all groups, as expected from the year-specific centering. White patients tend to show slightly positive relative efficiency scores across all periods, while other groups tend to fall marginally below their respective yearly averages. However, the differences are modest, indicating broad similarity in average efficiency across groups when conditioning on year and covariate structure. A more pronounced distinction emerges in the width of the prediction intervals. White patients exhibit the narrowest uncertainty intervals, indicating more homogeneous efficiency outcomes, whereas Asian, Black, and Hispanic patients have noticeably wider intervals, suggesting greater within-group heterogeneity in KPD transplant efficiency. It is important to emphasize that these wide group-conditional prediction intervals limit the strength of any claims about equality of efficiency across ethnic groups. By design, the conformal procedure is calibrated to achieve $95\%$ within-group coverage, and the interval width reflects both model uncertainty and the intrinsic variability of efficiency scores in a small, highly selected KPD population. As a result, the substantial overlap between intervals should not be interpreted as evidence that the allocation system is fair or that group-wise efficiency levels are identical. Rather, the table indicates that, given current sample sizes and noise levels, we do not detect large average differences in overall efficiency once priority, access, and outcome are aggregated and conditioned on covariates. In this sense, the DEA-based summaries are best viewed as diagnostic tools that complement the more pronounced criterion-specific disparities documented in Section~\ref{sec:3.3}, rather than as definitive proof of group-level fairness.

\label{sec:4.1_end}

\subsection{Analysis of Efficiency Distributions}
\label{sec:4.2}

To rigorously evaluate the statistical significance of the observed differences in efficiency distributions, we employed a kernel-based hypothesis-testing approach based on the Maximum Mean Discrepancy \citep[MMD,][]{Gretton2012}. We conducted both pairwise comparisons and "group-vs-rest" analyses across three cumulative time periods (2010-2013, 2010-2016, and 2010-2019) to assess how distributional differences evolve with sample size and temporal scope. For each comparison between groups $A$ and $B$, we defined the null hypothesis as $H_0: F_A = F_B$ and the alternative hypothesis as $H_1: F_A \neq F_B$, where $F_A$ and $F_B$ are the underlying distributions of relative efficiency scores.

The MMD between these distributions is defined as
\begin{equation*}
\text{MMD}^2[F_A, F_B] = \mathbb{E}_{\theta_A,\theta_A'}[K(\theta_A,\theta_A')] + \mathbb{E}_{\theta_B,\theta_B'}[K(\theta_B,\theta_B')] - 2\mathbb{E}_{\theta_A, \theta_B}[K(\theta_A,\theta_B)],
\end{equation*}
where $\theta_A, \theta_A' \stackrel{iid}{\sim} F_A$ and $\theta_B, \theta_B' \stackrel{iid}{\sim} F_B$ and $K(\cdot,\cdot)$ is a Laplacian kernel chosen to effectively detect variations in the distributions by measuring similarity between observations based on their absolute differences. Specifically, for efficiency scores $\theta_i$ and $\theta_j$, the Laplacian kernel is $K(\theta_i, \theta_j) = \exp\left(-\sigma \|\theta_i - \theta_j\|\right),
$ where $\sigma$ is set to the median Euclidean distance between points in the aggregate sample for each comparison. This kernel achieves maximum similarity (value of 1) when two observations are identical and decreases exponentially as their absolute difference increases. The Laplacian kernel is particularly well-suited for our context because it is sensitive to local distributional differences, making it effective at detecting variations in distributional spread and heterogeneity \citep{Gretton2012}. This choice aligns with our finding that White patients exhibit narrower prediction intervals (indicating more homogeneous outcomes) while Asian, Black, and Hispanic patients show wider intervals (suggesting greater within-group heterogeneity). The Laplacian kernel effectively captures these differences in distributional shape and spread, which are central to understanding equity in transplant outcomes. We determined statistical significance via permutation tests with 500 Monte Carlo iterations and applied Benjamini-Hochberg correction to control the false discovery rate across multiple comparisons.

\begin{table}[tbh]
\centering

\caption{MMD Test Results Across Time Periods. * indicates $p < 0.05$ after Benjamini-Hochberg correction.}
\label{tab:mmd_results}
\begin{tabular}{llccc}
\hline
Period & Comparison & MMD Statistic & $p$-value & Adjusted $p$-value \\
\hline
\multirow{10}{*}{\textbf{2010-2013}} 
 & Asian vs. Black & -0.0002 & 0.427 & 0.577 \\
 & Asian vs. Hispanic & -0.001 & 0.792 & 0.792 \\
 & Asian vs. White & -0.00002 & 0.319 & 0.577 \\
 & Black vs. Hispanic & 0.0004 &  0.154 & 0.533 \\
 & Black vs. White & -0.0001 &  0.481 & 0.577 \\
 & Hispanic vs. White & 0.0003 &  0.178 & 0.533 \\
 \cline{2-5}
 & Asian vs. Rest & -0.0003 & 0.457 & 0.457 \\
 & Black vs. Rest & -0.000008 & 0.351 & 0.457 \\
 & Hispanic vs. Rest & 0.0003 & 0.176 & 0.457 \\
 & White vs. Rest & -0.00002 & 0.377 &  0.457 \\
\hline
\multirow{10}{*}{\textbf{2010-2016}} 
 & Asian vs. Black & -0.0004 & 0.992 & 0.992 \\
 & Asian vs. Hispanic & -0.0002 & 0.637 & 0.802 \\
 & Asian vs. White & -0.0002 & 0.669 & 0.802 \\
 & Black vs. Hispanic & 0.00008 & 0.202 & 0.403 \\
 & Black vs. White & 0.00004 & 0.172 & 0.403 \\
 & Hispanic vs. White & 0.0004 & 0.004 & \ \ \ 0.024\ * \\
 \cline{2-5}
 & Asian vs. Rest & -0.0003 & 0.946 & 0.946 \\
 & Black vs. Rest & -0.00003 & 0.559 & 0.745 \\
 & Hispanic vs. Rest & 0.0002 & 0.048 & 0.096 \\
 & White vs. Rest & 0.0001 & 0.028 &  0.096 \\
\hline
\multirow{10}{*}{\textbf{2010-2019}} 
 & Asian vs. Black & -0.000002 & 0.355 & 0.563 \\
 & Asian vs. Hispanic & -0.000003 & 0.375 & 0.563 \\
 & Asian vs. White & -0.00008 & 0.928 & 0.928 \\
 & Black vs. Hispanic & -0.00004 & 0.910 & 0.928 \\
 & Black vs. White & 0.00007 & 0.020 & 0.120 \\
 & Hispanic vs. White & 0.00004 & 0.120 & 0.359 \\
 \cline{2-5}
 & Asian vs. Rest & -0.00006 & 0.677 & 0.677 \\
 & Black vs. Rest & 0.00003 & 0.080 & 0.160 \\
 & Hispanic vs. Rest & -0.000002 & 0.389 & 0.519 \\
 & White vs. Rest & 0.000052 & 0.006 & \ \ \ 0.024\ *  \\
\hline
\end{tabular}

\end{table}

The results in Table \ref{tab:mmd_results} reveal a consistent pattern of distributional differences that becomes more detectable as sample sizes increase across time periods. In the earliest period (2010-2013), no significant differences were detected between any group pairs, likely due to limited statistical power with smaller sample sizes. As the sample expanded to the 2010-2016 period, the Hispanic vs. White comparison became statistically significant (adjusted $p$-value = 0.024), suggesting these groups have detectably different efficiency distributions. With the largest sample (2010-2019), while the pairwise Hispanic vs. White comparison loses significance, the "group-versus-rest" analysis reveals that White patients show a significant difference from all other groups combined (adjusted $p$-value = 0.024). The "group-versus-rest" analyses provide critical insights into how individual groups differ from the broader transplant population. The pattern shows that White patients consistently exhibit distributional characteristics that distinguish them from the broader transplant population, with this difference becoming statistically significant in the largest sample. This finding provides statistical evidence supporting the descriptive findings from Section~\ref{sec:4.1} regarding differential within-group variability across ethnic groups.

At the same time, the general lack of statistically significant pairwise differences across many comparisons should not be over-interpreted. Failure to reject $H_0: F_A = F_B$ does not imply that the underlying efficiency distributions are equal or that allocation outcomes are fair across ethnic groups. Rather, it reflects the combined effect of sample size, substantial within-group variability, and the fact that efficiency aggregates multiple fairness criteria that already exhibit clear disparities when analyzed individually. Within these limitations, the MMD tests indicate only that we do not detect large distributional shifts in \emph{overall} efficiency between most group pairs. These findings suggest that while average efficiency scores are similar across groups after accounting for temporal trends, there are meaningful differences in the distributional properties of efficiency scores that may reflect disparities in access to optimal transplant outcomes within the kidney paired donation system. Taken together, the MMD and descriptive results highlight DEA’s role as a synthesis and monitoring tool rather than a standalone determinant of fairness.

\subsection{Discussion and Implication for Practice}
\label{sec:4.3}

Our DEA-based fairness analysis has several implications for kidney exchange practices. First, the efficiency score distributions shown in Figure \ref{fig:ridge} and Table \ref{tab:intervals} show larger variability for Asian, Black, and Hispanic patients compared to White patients, who exhibit more consistent outcomes. This narrower distribution for White patients could be attributed to better access to healthcare resources or more effective pre-transplant care \citep{Kucirka2012}. From a practical standpoint, the combination of centered efficiency scores and group-conditional prediction intervals provides a summary diagnostic for each ethnic group that goes beyond average outcomes: programs can monitor whether a group’s efficiency distribution remains systematically wider or more skewed than that of others, even when mean performance is comparable. Taken together, these findings suggest the need for tailored interventions to improve outcome consistency for underrepresented groups \citep{Gordon2014}.

Second, incorporating multiple efficiency measures, as in the DEA model \eqref{eq:dea_model}, can lead to a fairer kidney allocation system across ethnic groups. While recent developments in kidney paired donation programs, such as integrating deceased donors as nondirected donors and chain initiators \citep{Wang2021, Wang2022}, have improved matching opportunities, these initiatives primarily focus on enhancing overall efficiency. However, our DEA framework emphasizes the trade-offs between various fairness criteria. For example, reducing waitlist times for Asian patients might negatively impact graft survival, while prioritizing graft survival for White patients could increase their wait times. To address these trade-offs and improve allocation strategies, we recommend using the conditional DEA framework to assess the impact of policy changes on overall fairness across ethnic groups. In practice, this does not certify that a given allocation rule is fair, but it provides a structured way to quantify how proposed changes shift group-wise efficiency distributions while jointly accounting for multiple fairness-relevant criteria of the allocation process.

A concrete way to use our framework in practice is via policy simulation. Modern KPD platforms such as StanfordKPD \citep{stanfordkpd}, which support intelligent matching, prioritization of urgent patients, adjustable chain-length rules, and fine-grained immunologic constraints, allow programs to rerun historical pools or generate counterfactual match-runs under modified allocation policies. For example, a KPD program considering changes aimed at reducing long wait times among Asian and Black patients, such as increasing the priority weights for these groups in chain initiation or bridge-donor selection, can implement these adjustments within a platform like StanfordKPD and obtain the resulting set of simulated match outcomes. Using these simulated transplants, the program can then apply predictive models to estimate the priority, access, and outcome criteria scores for each match and compute conditional DEA scores via \eqref{eq:dea_model} under the proposed mechanism. Comparing the resulting efficiency distributions and their group-conditional prediction intervals to those under the status quo provides a principled diagnostic for assessing whether the policy improves group-wise fairness without inducing large efficiency losses or excessive variability for other groups. In this way, the conditional DEA scores serve as a common currency for evaluating competing allocation proposals that alter priority, access, and outcome in different ways. More broadly, the conditional DEA and Reference Frontier Mapping (RFM) framework introduced in Section~\ref{sec:uq-rfm} can be embedded into a routine monitoring pipeline. Programs can periodically reconstruct the RFM reference set, recompute group-specific conditional efficiency scores via \eqref{eq:dea_model}, and update group-conditional prediction intervals using the conformal procedure of Section~\ref{sec:uq-rfm} as new transplants accrue. Because covariates localize the production frontier to $\mathcal{T}_{Z_i}$, observed disparities in these diagnostics can be interpreted as differences in how effectively similar patients are treated rather than as artifacts of case-mix variation. This enables a feedback-driven workflow in which clinicians and policymakers (i) use standard clinical tools to design candidate changes to matching rules or eligibility criteria, (ii) screen those changes with DEA and RFM-based fairness diagnostics, and (iii) implement only policies that improve group-wise efficiency distributions while maintaining, at minimum, current outcomes for all other groups.

Despite these advantages, several practical barriers limit the immediate implementation of DEA-based fairness monitoring in real KPD programs. First, our framework requires detailed, high-quality covariate data to construct conditional frontiers and RFM reference sets; in practice, centers differ in how completely they record socioeconomic variables, post-transplant care, and follow-up outcomes, which can constrain the reliability of group-wise efficiency estimates for smaller subpopulations. Additionally, translating fairness diagnostics into policy requires agreement among clinicians, administrators, and regulators on acceptable trade-offs between efficiency and equity; interventions that narrow the efficiency distribution for one group may be perceived as limiting short-term matching flexibility or complicating existing allocation rules. Finally, routine monitoring can only be sustained if the resulting flags are interpretable and actionable; without clear governance about how to respond when a particular group's efficiency distribution degrades, there is a risk that fairness reports become merely procedural rather than a driver of meaningful change. These barriers suggest that, at least initially, our framework is most realistically deployed as an offline audit tool that is used periodically to evaluate the implications of proposed policy changes and to benchmark centers, prior to being integrated into real-time decision support as data infrastructure and institutional capacity improve.

\section{Conclusion}
\label{sec:5}
We propose a new method for evaluating fairness in kidney allocation using Data Envelopment Analysis (DEA), which can be broadly applied to assess fairness in complex healthcare systems. Our DEA framework considers multiple fairness criteria—priority, access, and outcome—providing a more comprehensive view of disparities compared to traditional single-metric approaches. The analysis shows that White patients tend to have more consistent outcomes, while Asian, Black, and Hispanic groups exhibit wider efficiency distributions. The advantage of handling multiple inputs and outputs without predefined weights in DEA is that it allows for an objective assessment of fairness. Our approach highlights where disparities emerge across different fairness criteria and offers a diagnostic framework for policymakers to evaluate how candidate policies might improve group-wise efficiency, while also making explicit the data, modeling, and governance infrastructure required for routine deployment. The code for reproducing the numerical results is available at \url{https://github.com/amofrad/Kidney-Exchange-DEA}.

Several research directions could build on this conditional DEA-based framework.  For example, dynamic conditional DEA models could capture how fairness evolves as KPD programs and matching algorithms change. Beyond healthcare, this approach has potential applications in evaluating fairness in diverse resource allocation contexts where resource scarcity and mutual compatibility constraints are present. For instance, in college admissions, the framework could evaluate fairness in student selection processes, accounting for factors such as academic performance, extracurricular activities, and institutional diversity goals. In job candidate selection, it could evaluate the fairness of hiring practices by simultaneously considering applicant qualifications, company needs, and diversity initiatives.

\begin{funding}
    This work was partially supported by NIH grant R01DK142026, NIH dkNET AI Pilot Program, a Merck Research Award, and a Hellman Fellowship Award.
\end{funding}


\bibliographystyle{imsart-nameyear} 
\bibliography{main}

\appendix

\newpage

\begin{center}
\large{\textbf{Supplementary Appendix}}
\end{center}

\noindent
This supplement provides additional methodological details, theoretical guarantees, and empirical validation for the proposed framework. We first present the group-conditional uncertainty quantification procedure under Reference Frontier Mapping (Section~S1), then provide detailed proofs of the main theorems (Section~S2), and finally conduct a comprehensive simulation study of our resampling method for conditional Data Envelopment Analysis (DEA) efficiency scores and additional variable-selection analyses (Section~S3).

\color{black}
\section*{S1. Group-Conditional Uncertainty Quantification under Reference Frontier Mapping}

\subsection*{S1.1 Construction of the Reference Set and Conditional Frontiers}

The Reference Frontier Mapping (RFM) framework partitions the data into a reference set $\mathcal{D}_\text{ref}$ and an evaluation set $\mathcal{D}_\text{eval}$ with the goal of using $\mathcal{D}_\text{ref}$ to estimate covariate-conditional frontiers that are informative about the boundary of the production possibility set, while reserving $\mathcal{D}_\text{eval}$ for uncertainty quantification via conformal prediction. To ensure that $\mathcal{D}_\text{ref}$ is both informative and diverse, we first select all patients who fall in the lower $5$th percentile of either input variable (waitlist duration or LKDPI) or the upper $95$th percentile of the output (graft lifespan). These observations are likely to lie near the production frontier and thus carry high informational value for boundary estimation. If this initial subset exceeds the desired reference set size ($10\%$ of the full dataset), we apply a multiplicative frontier-relevance scoring system to retain the most boundary-defining observations.

For each observation $i$, let $(X_i,Y_i)$ denote the $p$-dimensional input vector and $q$-dimensional output vector, respectively. We define dimension-specific scores that measure proximity to the efficient frontier. For the $p$ input dimensions, we set
$$
s_k^{(i)} \;=\; \bigl(X_{ki} - \min_j X_{kj} + 1\bigr)^{-1},
\qquad k = 1,\ldots,p,
$$
which assigns higher scores to observations with lower (more favorable) input values. For the $q$ output dimensions, we define
$$
s_{p+\ell}^{(i)} \;=\; \frac{Y_{\ell i} - \min_j Y_{\ell j}}{\max_j Y_{\ell j} - \min_j Y_{\ell j}},
\qquad \ell = 1,\ldots,q,
$$
which assigns higher scores to observations with higher output values. The overall frontier-relevance score for observation $i$ is then
$$
F^{(i)} \;=\; \prod_{k=1}^{p+q} s_k^{(i)},
$$
and we rank observations in the initial extreme subset by descending $F^{(i)}$ to select those most likely to define the production frontier boundaries. The remaining patients form the evaluation set $\mathcal{D}_\text{eval}$, and by construction $\mathcal{D}_\text{ref} \cap \mathcal{D}_\text{eval} = \varnothing$.

For each evaluation unit $i \in \mathcal{D}_\text{eval}$, we define a covariate-conditional production set using only reference units with positive weight under the kernel $K_h$ in (2) of the manuscript. Let $Z_i$ denote the vector of exogenous covariates for unit $i$ (e.g., education, UNOS region, citizenship). We set
$$
\mathcal{I}_i \;=\; \bigl\{\, j \in \mathcal{D}_\text{ref} : K_h(Z_j,Z_i) > 0 \,\bigr\}
$$
and define the covariate-conditional production set
$$
\mathcal{T}_{Z_i} = \left\{ 
(X, Y) \in \mathbb{R}_+^p \times \mathbb{R}_+^q \ \middle|\ 
X \geq \sum_{j \in \mathcal{I}_i} \lambda_j X_j,\ 
Y \leq \sum_{j \in \mathcal{I}_i} \lambda_j Y_j,\
\sum_{j \in \mathcal{I}_i} \lambda_j = 1,\ 
\lambda_j \geq 0
\right\},
$$
as in (1), but now constructed solely from $\mathcal{D}_\text{ref}$. This yields a locally adaptive frontier that reflects the structural and demographic context of KPD transplant patient $i$. The conditional efficiency score for unit $i$ is then defined by solving the conditional DEA optimization problem in (4); equivalently,
\[
\theta_i \;=\; \varphi_{\mathcal{T}_{Z_i}}(X_i, Y_i),
\]
where $\varphi_{\mathcal{T}_{Z_i}} : \mathbb{R}^p \times \mathbb{R}^q \to \mathbb{R}$ denotes the DEA operator induced by the covariate-conditional frontier $\mathcal{T}_{Z_i}$. This score captures the minimum proportional contraction of inputs and expansion of outputs needed to project unit $i$ onto its covariate-conditional frontier constructed from $\mathcal{D}_\text{ref}$. Once $\mathcal{D}_\text{ref}$ is fixed, the evaluation triplets $\{(X_i,Y_i,\theta_i)\}_{i \in \mathcal{D}_\text{eval}}$ are conditionally independent of the frontier construction. Hence, for conformal prediction we may treat them as conditionally i.i.d. draws from a data-generating process with fixed mapping $\varphi_{\mathcal{T}_{Z_i}}$, resolving the dependence issue that arises when the frontier is re-estimated on the full dataset.

\vspace{0.4cm}

\subsection*{S1.2 Group-Conditional Conformal Prediction}

We now describe the group-conditional conformal prediction procedure used to obtain uncertainty guarantees for the efficiency scores $\theta_i$ under RFM. The construction follows \citet{Gibbs2023}, specialized to our setting, where the primary covariate of interest is ethnic group membership. Let $\mathcal{G}$ denote the set of groups, and let $(X_i,Y_i,Z_i)$ be i.i.d. samples from $\mathbb{P}_{X,Y,Z}$ with associated conditional efficiency scores $\theta_i = \varphi_{\mathcal{T}_{Z_i}}(X_i,Y_i)$ computed as in Section~S1.1. To quantify uncertainty in a group-aware manner and ensure group-conditional coverage, we define the function class
\begin{equation}
\label{eq:function_class}
\mathcal{F} = \left\{f(x,y) = \sum_{g \in \mathcal{G}} \beta_g \,\mathbb{I}\{ (x,y) \in g\} \mid \beta_g \in \mathbb{R}, \ \forall g \in \mathcal{G} \right\},
\end{equation}
where $\mathbb{I}\bigl\{(x,y)\in g\bigr\}$ is the indicator of group membership for the input–output pair $(x,y)$ and $\beta_g$ are group-specific coefficients. In our application, group membership is determined by patient ethnicity, so that $g$ identifies the ethnic group and $f(x,y)$ is a piecewise-constant function across groups. We use $\mathcal{F}$ to model the conditional mean efficiency score via
$$
\hat{\theta}(X_i,Y_i) \;=\; \sum_{g \in \mathcal{G}} \beta_g \,\mathbb{I}\{(X_i,Y_i)\in g\},
$$
and define the conformity score
$$
\mathcal{S}\bigl((X_i,Y_i),\theta_i\bigr) \;=\; \bigl|\theta_i - \hat{\theta}(X_i,Y_i)\bigr|.
$$
Intuitively, $\mathcal{S}$ measures how far an observed efficiency score deviates from its group-specific mean, and forms the basis for our group-conditional calibration. Following \citet{Gibbs2023}, the conformal calibration step is formulated through an optimization problem over dual variables $\eta \in \mathbb{R}^{n+1}$:
\begin{align}
\label{eq:opt_problem}
\max_{\eta \in \mathbb{R}^{n+1}} \ & \sum_{i=1}^n \eta_i \mathcal{S}_i + \eta_{n+1} \mathcal{S} - \mathcal{R}^*(\eta),  \\
\text{subject to } & -\alpha \leq \eta_i \leq 1-\alpha, \nonumber
\end{align}
where $\mathcal{S}_i = \mathcal{S}((X_i,Y_i),\theta_i)$ for $i=1,\dots,n$ and $\mathcal{S} = \mathcal{S}((X_{n+1},Y_{n+1}),\theta)$ is the conformity score associated with a candidate value $\theta$ for the $(n+1)$-th observation. The term $\mathcal{R}^*(\eta)$ is the convex conjugate of a quadratic regularization over $\mathcal{F}$:
\begin{equation}
\label{eq:regularization}
\mathcal{R}^*(\eta) = -\min_{h \in \mathcal{F}} \left\{(n+1)\lambda \|h\|_2^2 - \sum_{i=1}^{n+1} \eta_i h(X_i, Y_i) \right\},
\end{equation}
with regularization parameter $\lambda > 0$. The constraints $-\alpha \leq \eta_i \leq 1-\alpha$ control the influence of individual samples and are crucial for achieving the desired $(1-\alpha)$ coverage level within each group. Solving \eqref{eq:opt_problem} for each candidate $\mathcal{S}$ yields an optimal dual coordinate $\eta_{n+1}^{\mathcal{S}}$, which plays the role of a calibrated, group-aware conformity threshold. We then define the randomized prediction set
\begin{equation}
\label{eq:prediction_set}
\hat{C}_\text{rand.}(X_{n+1}, Y_{n+1}) = \{\theta: \eta_{n+1}^{\mathcal{S}((X_{n+1},Y_{n+1}),\theta)} < U\},
\end{equation}
where $U \sim \text{Unif}([-\alpha, 1 - \alpha])$ is drawn independently, ensuring exact finite-sample coverage.

Under the RFM framework, once the reference set $\mathcal{D}_\text{ref}$ is fixed, the evaluation samples ${(X_i,Y_i,Z_i)}$ and their corresponding scores ${\theta_i}$ are conditionally i.i.d., so the validity of the construction in \eqref{eq:prediction_set} follows directly and yields the group-conditional coverage guarantee stated in Theorem~1:
\[
\mathbb{P}\left(\theta_{n+1} \in \hat{C}_\text{rand.}(X_{n+1}, Y_{n+1}) \mid (X_{n+1}, Y_{n+1}) \in g\right) = 1 - \alpha, \quad \forall g \in \mathcal{G}.
\]
A detailed proof of this result, adapted to the RFM setting, is given in Section~S2.1.

\color{black}
\section*{S2. Proofs of Theorems}

\subsection*{S2.1 Proof of Theorem 1}
\label{sec:thm1}

\begin{proof}
We follow the approach and structure presented in \citet{Gibbs2023} to prove the group-conditional coverage property of our randomized prediction set under the Reference Frontier Mapping (RFM) framework.

\vspace{0.3cm}

We begin by recalling the definition of our randomized prediction set
\begin{equation*}
\hat{C}_\text{rand.}(X_{n+1}, Y_{n+1}) = \{\theta: \eta_{n+1}^{\mathcal{S}((X_{n+1},Y_{n+1}),\theta)} < U\},
\end{equation*}
where $(X_{n+1}, Y_{n+1})$ is the input-output pair for a new patient, $\theta$ is the DEA efficiency score, and $U \sim \text{Unif}([-\alpha, 1-\alpha])$ is drawn independently. We define conformity score $\mathcal{S}((X,Y),\theta) = |\theta - \hat{\theta}(X,Y)|$, where $\hat{\theta}(X,Y) = \sum_{g \in \mathcal{G}} \beta_g \mathbb{I}\{(X,Y) \in g\}$ estimates the conditional mean efficiency score.

Under the RFM framework, the evaluation samples $\{(X_i, Y_i, Z_i)\}_{i=1}^n$ are conditionally independent given the fixed reference set $\mathcal{D}_\text{ref}$. For each evaluation unit $i$, the conditional efficiency score is computed as $\theta_i = \varphi_{\mathcal{T}_{Z_i}}(X_i, Y_i)$, where $\varphi_{\mathcal{T}_{Z_i}} : \mathbb{R}^p \times \mathbb{R}^q \to \mathbb{R}$ denotes the conditional DEA operator defined by the covariate-conditional production set $\mathcal{T}_{Z_i}$ constructed from $\mathcal{D}_\text{ref}$ via kernel smoothing. Similarly, for the new sample $(X_{n+1}, Y_{n+1}, Z_{n+1}) \sim \mathbb{P}_{X,Y,Z}$, the efficiency score is defined as $\theta_{n+1} = \varphi_{\mathcal{T}_{Z_{n+1}}}(X_{n+1}, Y_{n+1})$ using the same reference frontier.

Recall our function class $\mathcal{F}$ defined as
\begin{equation*}
\mathcal{F} = \left\{f(x,y) = \sum_{g \in \mathcal{G}} \beta_g \mathbb{I}\{ (x,y) \in g\} \mid \beta_g \in \mathbb{R}, \forall g \in \mathcal{G} \right\},
\end{equation*}
where $\mathbb{I}\{ (x,y) \in g\}$ is the indicator function for group membership of the input-output pair and $\beta_g$ are the coefficients for each group $g \in \mathcal{G}$.

The values of $\eta_{n+1}^\mathcal{S}$ are derived from the following optimization problem:
\begin{align}
\label{eq:optimization_problem}
\max_{\eta \in \mathbb{R}^{n+1}} & \sum_{i=1}^n \eta_i \mathcal{S}_i + \eta_{n+1} \mathcal{S} - \mathcal{R}^*(\eta)  \\
\text{subject to } & -\alpha \leq \eta_i \leq 1-\alpha, \nonumber
\end{align}
where $\mathcal{R}^*(\eta)$ is the convex conjugate of the regularization term:
\begin{equation}
\label{eq:regularization1}
\mathcal{R}^*(\eta) = -\min_{h \in \mathcal{F}} \left\{(n+1)\lambda \|h\|_2^2 - \sum_{i=1}^{n+1} \eta_i h(X_i, Y_i) \right\},
\end{equation}
with $\lambda > 0$ as the regularization parameter.

Let $\mathbb{P}_g$ denote the conditional probability given $(X_{n+1}, Y_{n+1}) \in g$. Our goal is to demonstrate that:
\begin{equation*}
\mathbb{P}_g(\theta_{n+1} \in \hat{C}_{\text{rand.}}(X_{n+1}, Y_{n+1})) = 1 - \alpha, \quad \forall g \in \mathcal{G}.
\end{equation*}

\noindent By the definition of our prediction set, we can rewrite this probability as
\begin{equation*}
\mathbb{P}_g(\theta_{n+1} \in \hat{C}_{\text{rand.}}(X_{n+1}, Y_{n+1})) = \mathbb{P}_g(\eta_{n+1}^{\mathcal{S}((X_{n+1},Y_{n+1}),\theta_{n+1})} < U).
\end{equation*}

\noindent Since $\theta_{n+1} = \varphi_{\mathcal{T}_{Z_{n+1}}}(X_{n+1}, Y_{n+1})$ is computed relative to the fixed reference frontier $\mathcal{D}_\text{ref}$, the triplets $\{(X_i, Y_i, \theta_i)\}_{i=1}^{n+1}$ are conditionally independent given the fixed frontier. This conditional independence property ensures that the conformal prediction framework can be applied directly.

From the optimization problem \eqref{eq:optimization_problem}, the complementary slackness conditions imply
\begin{equation}
\label{eq:comp_slack}
\eta_{n+1}^\mathcal{S} = \begin{cases}
-\alpha & \text{if } \mathcal{S} < \hat{h}_\mathcal{S}(X_{n+1}, Y_{n+1}) \\
[-\alpha, 1-\alpha] & \text{if } \mathcal{S} = \hat{h}_\mathcal{S}(X_{n+1}, Y_{n+1}) \\
1-\alpha & \text{if } \mathcal{S} > \hat{h}_\mathcal{S}(X_{n+1}, Y_{n+1})
\end{cases}
\end{equation}
where $\hat{h}_\mathcal{S}$ is the fitted quantile function, which estimates the $\alpha$-quantile of the conformity score distribution conditional on the input-output pairs. It is related to the optimization problem through the function class as expressed in \eqref{eq:regularization1}.

Given that $U \sim \text{Unif}([-\alpha, 1-\alpha])$ and applying the conditions from \eqref{eq:comp_slack}, we have:

\small{\begin{equation}
\label{eq:prob_calculation}
\mathbb{P}_g(\eta_{n+1}^{\mathcal{S}((X_{n+1},Y_{n+1}),\theta_{n+1})} < U) = 
\begin{cases}
0 & \hspace{-0.2em}\text{if } \mathcal{S}((X_{n+1},Y_{n+1}),\theta_{n+1}) < \hat{h}_\mathcal{S}(X_{n+1}, Y_{n+1}) \\
1 - \alpha - \eta_{n+1}^\mathcal{S} & \hspace{-0.2em}\text{if } \mathcal{S}((X_{n+1},Y_{n+1}),\theta_{n+1}) = \hat{h}_\mathcal{S}(X_{n+1}, Y_{n+1}) \\
1 & \hspace{-0.2em}\text{if } \mathcal{S}((X_{n+1},Y_{n+1}),\theta_{n+1}) > \hat{h}_\mathcal{S}(X_{n+1}, Y_{n+1})
\end{cases}
\end{equation}}

We note that $\theta_{n+1} = \varphi_{\mathcal{T}_{Z_{n+1}}}(X_{n+1}, Y_{n+1})$ is drawn from a continuous distribution, as it is derived from the conditional DEA model with continuous inputs and outputs relative to a smooth kernel-weighted frontier. Additionally, $\hat{h}_\mathcal{S}$ is a continuous function. Consequently, the probability of the event $\mathcal{S}((X_{n+1},Y_{n+1}),\theta_{n+1}) = \hat{h}_\mathcal{S}(X_{n+1}, Y_{n+1})$ is zero. Given this continuity, we can simplify \eqref{eq:prob_calculation} to
\begin{equation}
\label{eq:final_prob}
\mathbb{P}_g(\theta_{n+1} \in \hat{C}_{\text{rand.}}(X_{n+1}, Y_{n+1})) = \mathbb{P}_g(\mathcal{S}((X_{n+1},Y_{n+1}),\theta_{n+1}) > \hat{h}_\mathcal{S}(X_{n+1}, Y_{n+1})).
\end{equation}

\noindent By the properties of quantile regression and the conditional independence established through RFM, we know that
\begin{equation}
\label{eq:quantile_prop}
\mathbb{P}_g(\mathcal{S}((X_{n+1},Y_{n+1}),\theta_{n+1}) > \hat{h}_\mathcal{S}(X_{n+1}, Y_{n+1})) = 1 - \alpha
\end{equation}
since $\hat{h}_\mathcal{S}(X_{n+1}, Y_{n+1})$ is the $\alpha$-quantile of the distribution of $\mathcal{S}((X_{n+1},Y_{n+1}),\theta_{n+1})$ conditional on $(X_{n+1}, Y_{n+1}) \in g$.

Combining \eqref{eq:final_prob} and \eqref{eq:quantile_prop}, we conclude that:
\begin{equation*}
\mathbb{P}_g(\theta_{n+1} \in \hat{C}_{\text{rand.}}(X_{n+1}, Y_{n+1})) = 1 - \alpha, \quad \forall g \in \mathcal{G},
\end{equation*}
which completes the proof.
\end{proof}

\subsection*{S2.2 Proof of Theorem 2}

\begin{proof}
We aim to prove that the bias of the resampled estimator is no larger than that of the imbalanced estimator for each group $g$, provided the resampled sample size $n'$ satisfies the stated lower bound.

\vspace{0.3cm}

Let $(X_i, Y_i, Z_i)$ be the input–output covariate combination for patient $i$. Under the conditional DEA framework, the efficiency score $\theta_i(Z_i)$ for patient $i$ with covariate profile $Z_i$ is obtained by solving

\begin{equation*}
\begin{aligned}
\min_{\theta,\,\{\lambda_j\}} \ & \theta \\
\text{s.t.}\quad & \theta\,X_i \;\ge\; \sum_{j \in \mathcal{I}_i} \lambda_j\,X_j, \qquad
\frac{1}{\theta}\,Y_i \;\le\; \sum_{j \in \mathcal{I}_i} \lambda_j\,Y_j, \\
& \sum_{j \in \mathcal{I}_i} \lambda_j = 1,\quad \lambda_j \,\ge\,0,
\end{aligned}
\end{equation*}
where $\mathcal{I}_i = \{ j \in \{1,\dots,n\} : K_h(Z_j, Z_i) > 0 \}$ and $K_h(\cdot,\cdot)$ is the smoothing kernel in (2) of the manuscript that measures similarity between covariate profiles. This problem defines a covariate‐conditional production set $\mathcal{T}_{Z_i}\subset\mathbb{R}_+^{p+q}$.

Let $\mathbb{P}_{X,Y,Z}$ denote the true joint distribution of $(X,Y,Z)$.  Let $\mathbb{P}_n$ be the empirical measure based on the imbalanced sample of size $n$, and $\mathbb{P}_{n'}$ be the empirical measure based on the resampled dataset of size $n'\le n$.  For each group $g$, define
$$
\theta_g^* \;:=\; \mathbb{E}_{(X,Y,Z)\sim P_g}\bigl[\theta(X,Y\mid Z)\bigr],
$$
where $P_g$ is the conditional distribution of $(X,Y,Z)$ given group $g$.  Denote by $\hat{\theta}_g$ the empirical mean of the conditional DEA scores under $\mathbb{P}_n$, and by $\hat{\theta}_g'$ the empirical mean under $\mathbb{P}_{n'}$. Then
$$
\mathrm{Bias}\bigl(\hat{\theta}_g\bigr) \;=\; \mathbb{E}_{\mathbb{P}_n}\bigl[\hat{\theta}_g\bigr] \;-\; \theta_g^*, 
\qquad
\mathrm{Bias}\bigl(\hat{\theta}_g'\bigr) \;=\; \mathbb{E}_{\mathbb{P}_{n'}}\bigl[\hat{\theta}_g'\bigr] \;-\; \theta_g^*.
$$
Let $\mathcal{T}_Z\subset\mathbb{R}_+^{p+q}$ be the true covariate‐conditional production set for each covariate profile $Z$.  Let $\hat{\mathcal{T}}_{Z,n}$ and $\hat{\mathcal{T}}_{Z,n'}$ be the estimated covariate‐conditional sets arising from the imbalanced and resampled samples, respectively.  By assumption, for each fixed $Z$, $\mathcal{T}_Z$ is compact with a $C^2$-smooth boundary, and the kernel $K_h$ is Lipschitz‐continuous in $Z$.  Standard results on nonparametric frontier estimation \citep{buadin2012measure} imply that, with probability one, $\hat{\mathcal{T}}_{Z,n} \;\subseteq\; \mathcal{T}_Z 
\quad\text{and}\quad \hat{\mathcal{T}}_{Z,n'} \;\subseteq\; \mathcal{T}_Z,$ meaning both estimators underestimate the true production set. 

We quantify the approximation error via the Hausdorff distance.  For any two sets $A,B\subset\mathbb{R}_+^{p+q}$, let
\begin{equation*}
d_H(A,B) = \max\{\sup_{a\in A}\inf_{b\in B}d(a,b), \ \sup_{b\in B}\inf_{a\in A}d(a,b)\},
\end{equation*}
where $d(a,b)$ is the distance between points $a$ and $b$. Because the resampling procedure balances group representations more evenly (especially within local neighborhoods defined by $K_h$), it follows that
$$
\mathbb{E}\bigl[d_H\bigl(\hat{\mathcal{T}}_{Z,n'},\,\mathcal{T}_Z\bigr)\bigr]
\;\le\;
\mathbb{E}\bigl[d_H\bigl(\hat{\mathcal{T}}_{Z,n},\,\mathcal{T}_Z\bigr)\bigr],
$$
for all covariate values $Z$. The Hausdorff distance measures the worst‐case distance between two sets, with a smaller distance indicating closer proximity between all points in the sets. Thus, the inequality $\mathbb{E}[d_H(\hat{\mathcal{T}}_{Z,n'}, \mathcal{T}_Z)] \le \mathbb{E}[d_H(\hat{\mathcal{T}}_{Z,n}, \mathcal{T}_Z)]$ implies that $\hat{\mathcal{T}}_{Z,n'}$ is, on average, a better approximation of $\mathcal{T}_Z$ than $\hat{\mathcal{T}}_{Z,n}$, particularly in regions where minority groups are concentrated. This improved approximation results from the resampling procedure's more balanced representation across groups within each local neighborhood defined by the kernel weights. Since DEA efficiency scores quantify the proportional contraction needed to reach the frontier, a closer approximation of $\mathcal{T}_Z$ leads to more accurate efficiency scores and smaller bias in the group‐level efficiency estimates $\hat{\theta}_g'$.

Under the conditional DEA framework with covariate dimension $r$, the convergence rate is affected by the effective sample size $nh^r$, where $h$ is the bandwidth. With optimal bandwidth $h = n^{-1/(r+4)}$, the conditional DEA estimators achieve a convergence rate of $n^{-4/((r+4)(p+q))}$ \citep{buadin2012measure}. Concretely,
$$
\bigl|\hat{\theta}_g - \theta_g^*\bigr|
\;=\;O_p\bigl(n^{-4/((r+4)(p+q))}\bigr),
\qquad
\bigl|\hat{\theta}_g' - \theta_g^*\bigr|
\;=\;O_p\bigl(n'^{-4/((r+4)(p+q))}\bigr).
$$
Define
$$
K \;:=\; \sup_{g\in\mathcal{G}}
\Bigl\{
\limsup_{n\to\infty}\,
n^{4/((r+4)(p+q))} 
\,\mathbb{E}\bigl[\bigl|\hat{\theta}_g - \theta_g^*\bigr|\bigr]
\Bigr\},
$$
which is finite under our smoothness assumptions.  Then for large $n$,
$$
\mathbb{E}\bigl[\bigl|\hat{\theta}_g' - \theta_g^*\bigr|\bigr]
\;\approx\;
K\,n'^{-4/((r+4)(p+q))}.
$$
Hence, to guarantee $\bigl|\mathrm{Bias}(\hat{\theta}_g')\bigr|\le \delta$, it suffices that
$$
n' \;\ge\;\Bigl(\tfrac{K}{\delta}\Bigr)^{\frac{(r+4)(p+q)}{4}}.
$$
Moreover, to ensure every group $g$ appears sufficiently often in the resampled set, we require
$$
n' \;\ge\;\min_{g\in\mathcal{G}}\frac{1}{p_g},
$$
where $p_g$ is the population‐level proportion of group $g$.  Combining these two lower bounds on $n'$ yields
$$
n' \;\ge\; \max\Bigl\{\,\bigl(K/\delta\bigr)^{\frac{(r+4)(p+q)}{4}}, \;\min_{g\in\mathcal{G}}\tfrac{1}{p_g}\Bigr\}.
$$
Under this condition, for each group $g$,
$$
\bigl|\mathrm{Bias}(\hat{\theta}_g')\bigr|
\;=\;\Bigl|\mathbb{E}_{\mathbb{P}_{n'}}[\hat{\theta}_g'] \;-\;\theta_g^*\Bigr|
\;\le\;
\Bigl|\mathbb{E}_{\mathbb{P}_n}[\hat{\theta}_g] \;-\;\theta_g^*\Bigr|
\;=\;
\bigl|\mathrm{Bias}(\hat{\theta}_g)\bigr|.
$$

This completes the proof that the resampled estimator's bias does not exceed that of the original imbalanced estimator for any group $g$.
\end{proof}

\section*{S2. Simulation Study: Impact of Resampling on Efficiency Scores}

\subsection*{S2.1 Synthetic Data Generation}

Our simulation procedure generated synthetic datasets ($n$ = 1000) that mimicked the structure and characteristics of our original data. Let $Z_g = (X_{1g}, X_{2g}, Y_{1g})^T$ represent the vector of variables for ethnic group $g$, where $X_{1g}$, $X_{2g}$, and $Y_{1g}$ denote waitlist duration, LKDPI, and graft lifespan, respectively. We modeled $Z_g$ using a multivariate normal distribution $Z_g \sim \mathcal{N}(\mu_g, \Sigma_g)$, where $\mu_g = ({\mu_{1g}, \mu_{2g}, \mu_{3g}})^T$ is the mean vector and $\Sigma_g$ is the covariance matrix for group $g \in \mathcal{G}$. Using empirical means and covariance matrices from our original dataset, we generated synthetic data maintaining the original ethnic proportions (Asian = 7.3\%, Black = 15.8\%, Hispanic = 14.4\%, White = 62.5\%). This approach ensured that our simulated data closely resembled real-world data in both individual-variable distributions and inter-variable relationships within each ethnic group, while preserving the original ethnic composition.

\subsection*{S2.2 Simulation Procedure and Results}

Our simulation process employed the conditional DEA framework with Reference Frontier Mapping. For each iteration, we generated a synthetic dataset of 1000 observations, applied RFM to separate reference and evaluation sets, computed conditional DEA efficiency scores for the original (imbalanced) dataset, applied our resampling method to create a balanced dataset matching ESRD prevalence rates, and computed conditional DEA efficiency scores for the resampled dataset. We repeated this process 100 times to ensure robust results while accounting for the computational complexity of the conditional DEA approach. Table \ref{tab:sim_summary} summarizes the results.

\begin{table}[H]
\centering
\caption{Summary Statistics of Relative Efficiency Scores from Simulation}
\label{tab:sim_summary}
\begin{tabular}{lcccccc}
\hline
\multirow{2}{*}{Group} & \multicolumn{3}{c}{Imbalanced} & \multicolumn{3}{c}{Resampled (to ESRD rates)} \\
\cline{2-4} \cline{5-7}
& Mean & SD & 95\% CI  & Mean & SD & 95\% CI  \\
\hline
Asian    & -0.00153 & 0.0981 & [-0.114, 0.253] & -0.000121 & 0.0813 & [-0.0954, 0.000] \\
Black    & -0.00139 & 0.103  & [-0.112, 0.226] & -0.000238 & 0.0814 & [-0.0953, 0.000] \\
Hispanic & -0.000842 & 0.104 & [-0.119, 0.253] & -0.000580 & 0.0786 & [-0.0953, 0.000] \\
White    & 0.000721 & 0.110  & [-0.113, 0.300] & 0.000397  & 0.0874 & [-0.0954, 0.000] \\
\hline
\end{tabular}
\end{table}

The simulation results reveal important patterns in how our year-specific ESRD-based resampling method affects relative efficiency score estimates across different ethnic groups under the conditional DEA framework. All minority groups show slight improvements in mean efficiency scores after resampling, with reduced variability as indicated by smaller standard deviations. The resampling procedure appears to reduce the spread of efficiency distributions, leading to more consistent outcomes across all ethnic groups. This suggests that accounting for proper demographic representation through ESRD prevalence rates yields more stable and representative efficiency estimates.

\subsection*{S2.3 Sensitivity Analysis}

\begin{figure}[H]
\centering
\includegraphics[width=0.9\textwidth]{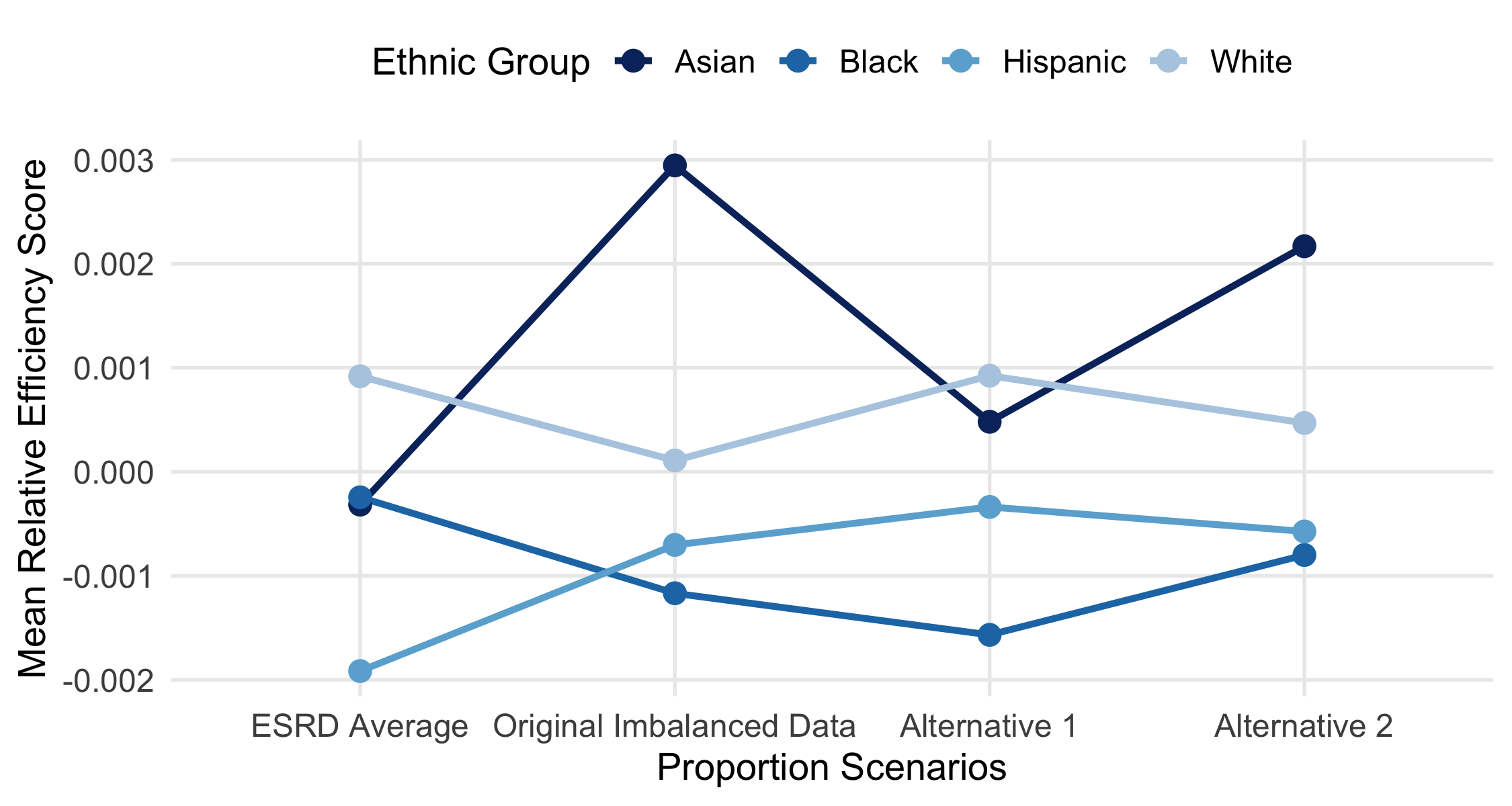}
\caption{Relative Efficiency Scores Across Different Proportion Scenarios}
\label{fig:sensitivity_comparison}
\end{figure}

To validate our findings and assess the robustness of the results, we conducted a sensitivity analysis across four population proportion scenarios (Table \ref{tab:population_proportions}). For each scenario, we generated synthetic datasets using ESRD average proportions as the baseline, calculated conditional DEA efficiencies using RFM, resampled the data according to alternative proportion scenarios, and computed corresponding efficiency scores. We repeated this process 100 times for each scenario. The results are visualized in Figure \ref{fig:sensitivity_comparison} and summarized in Table \ref{tab:sensitivity_comp2}.

\begin{table}[tbh]
\centering
\caption{Population Proportions Scenarios}
\label{tab:population_proportions}
\begin{tabular}{lcccc}
\hline
Proportion Type & Asian & Black & Hispanic & White \\
\hline
ESRD Average & 5.4\% & 30.0\% & 19.1\% & 45.5\% \\
Original Imbalanced Data & 7.3\%  & 15.8\%  & 14.4\%  & 62.5\%  \\
Alternative 1 & 10\%  & 20\%  & 30\%  & 40\%  \\
Alternative 2 & 20\%  & 40\%  & 30\%  & 10\%  \\
\hline
\end{tabular}
\end{table}

\begin{table}[H]
\centering
\caption{Mean Relative Efficiency Scores Across Different Proportion Scenarios}
\label{tab:sensitivity_comp2}
\begin{tabular}{lcccc}
\hline
Proportion Type & Asian & Black & Hispanic & White \\
\hline
ESRD Average &           -0.000316 & -0.000245 & -0.00192 & 0.000919 \\
Original Imbalanced Data & 0.00295   & -0.00117  & -0.000703 & 0.000109 \\
Alternative 1  &          0.000481  & -0.00157  & -0.000338 & 0.000925 \\
Alternative 2   &         0.00217   & -0.000800 & -0.000574 & 0.000469 \\
\hline
\end{tabular}
\end{table}

The sensitivity analysis provides compelling evidence for the necessity of resampling to ensure data accurately reflects ESRD demographics when assessing fairness in kidney allocation systems. Asian patients show positive efficiency scores in all scenarios except the ESRD Average baseline, suggesting that demographic imbalances may lead to overestimation of their relative performance. Black patients consistently show slightly negative efficiency scores across most scenarios, with the least negative performance under the Original Imbalanced Data scenario where they are underrepresented. Hispanic patients demonstrate relatively stable efficiency scores across scenarios, though slightly negative in the ESRD Average scenario. White patients show consistently positive efficiency scores across all scenarios, with the highest performance under Alternative 1 where they maintain substantial representation.

The variations observed across Alternative 1 and Alternative 2 reinforce the importance of accurate demographic representation in conditional DEA analysis. The conditional framework, combined with RFM, appears to yield more stable estimates than traditional DEA approaches, as evidenced by smaller differences between scenarios and narrower confidence intervals.

\subsection*{S2.4 Quantifying Bias}

To quantify bias introduced by different population proportions in our conditional DEA framework, we calculated the difference between mean relative efficiency scores under each alternative scenario and the ESRD average proportions. We define the bias for each group and proportion scenario as $\text{Bias}_{g,p}(\theta) = \bar{\theta}_{g,p} - \bar{\theta}_{g,\text{ESRD}}$, where $\bar{\theta}_{g,p}$ and $\bar{\theta}_{g,\text{ESRD}}$ denote mean relative efficiency scores for ethnic group $g$ under proportion scenario $p$ and ESRD average proportions, respectively. We computed a weighted average bias for each scenario as $\text{Weighted Bias}_p (\theta) = \sum_{g} w_{g,p} \cdot \text{Bias}_{g,p}(\theta)$, where $w_{g,p}$ is the proportion of group $g$ in scenario $p$. Table \ref{tab:bias_results} presents the calculated biases.

\begin{table}[H]
\centering
\caption{Bias in Relative Efficiency Scores Across Different Proportion Scenarios}
\label{tab:bias_results}
\begin{tabular}{lcccc}
\hline
Group & ESRD Average & Original Imbalanced Data & Alternative 1 & Alternative 2 \\
\hline
Asian    & - & 0.00326 & 0.000797 & 0.00248  \\
Black    & - & -0.000923  & -0.00132   &  -0.000555 \\
Hispanic & - & 0.00121   & 0.00158 & 0.00134   \\
White    & - &  -0.000810  & 0.00000587 & -0.000449 \\
\hline
Weighted Bias & - & -0.000261 &  0.000291 & 0.000633 \\
\hline
\end{tabular}
\end{table}

The bias analysis reinforces the importance of using accurate demographic proportions in fairness assessments of kidney allocation systems. Under the original imbalanced data proportions, Asian and Hispanic patients show positive bias, while Black and White patients show negative bias, with the overall weighted bias being slightly negative. Alternative 1 introduces primarily positive biases across most groups, while Alternative 2 shows mixed effects. The weighted bias calculations reveal that Alternative 2 presents the largest overall positive bias, followed by Alternative 1, while the original imbalanced data proportions introduce a small negative overall bias.

These findings highlight the sensitivity of conditional DEA efficiency measures to demographic composition, even after controlling for covariate differences via kernel-weighted local frontiers. The non-uniform response of different ethnic groups to changing proportions suggests that fairness in kidney allocation requires careful attention to both the methodological framework and the demographic representation in the dataset. While the biases are relatively small under the conditional DEA framework, they demonstrate consistent patterns that could influence fairness conclusions, particularly when the analysis is extended to larger datasets or different allocation systems. These results emphasize the importance of using accurate sample proportions in fairness assessments and highlight the potential pitfalls of applying fairness metrics developed in one demographic context to populations with different compositions.

\section*{S3. Variable Selection Testing}

\subsection*{S3.1 Variable Selection for Mediation Analysis}

To determine which variables should be included as mediators versus covariates in our Priority Fairness mediation analysis, we implemented a structured two-stage hypothesis testing procedure based on the mediation testing framework of \citet{mackinnon2007mediationtests}. This approach enables systematic identification of variables that lie along causal pathways between ethnicity and waitlist duration (mediators) versus those that directly influence outcomes independent of group membership (covariates).

\subsubsection*{Stage 1: Association with Outcome Variable}

In the first stage, we assessed the association between each candidate variable $M_j \in M$ and the outcome $X_1$ (waitlist duration) using likelihood ratio tests within a generalized linear modeling framework. Each test compared a full model including $M_j$ to a reduced model excluding it, thereby evaluating the incremental explanatory power of $M_j$ conditional on the other variables.

Let $G$ denote the group indicator for ethnicity, and let $M = (M_1, M_2, \dots, M_k)$ denote the vector of $k$ candidate mediators and covariates. For each variable $M_j \in M$, we define the full model as:
$$
X_1 = \beta_0 + \beta_G G + \sum_{\substack{l=1}}^{k} \beta_l M_l + \varepsilon,
$$
and the reduced model (excluding $M_j$) as:
$$
X_1 = \beta_0^{(-j)} + \beta_G^{(-j)} G + \sum_{\substack{l=1 \\ l \neq j}}^{k} \beta_l^{(-j)} M_l + \varepsilon^{(-j)}.
$$

We then compute the likelihood ratio statistic for variable $M_j$ as:
$$
\Lambda_j = -2 \left( \log L_{\text{reduced}}^{(j)} - \log L_{\text{full}}^{(j)} \right),
$$
which is asymptotically distributed as $\chi^2_{d_j}$, where $d_j$ is the number of degrees of freedom associated with $M_j$. This test quantifies the additional explanatory power of $M_j$ for explaining $X_1$, conditional on the presence of all other covariates.

\begin{table}[H]
  \centering
  \caption{Stage 1: Association Between Candidate Variables and Waitlist Duration}
  \label{tab:stage1_mediator_response}
  \setlength{\tabcolsep}{4pt}
  \renewcommand{\arraystretch}{1.1}
  \begin{tabular}{@{}lcccc@{}}
    \toprule
    \textbf{Variable ($M_j$)} & \textbf{LR $\chi^2$} & \textbf{df ($d_j$)} 
      & \textbf{p-value} & \textbf{BH-adjusted p-value} \\
    \midrule
    REGION               & 47.34  & 10 & $8.17\times10^{-7}$$^{***}$ & $2.12\times10^{-6}$$^{***}$ \\
    EDUCATION            & 31.13  & 7  & $5.87\times10^{-5}$$^{***}$ & $1.09\times10^{-4}$$^{***}$ \\
    CITIZENSHIP          & 30.14  & 3  & $1.29\times10^{-6}$$^{***}$ & $2.80\times10^{-6}$$^{***}$ \\
    PRI\_PAYMENT\_TCR\_KI &  4.96  & 7  & $0.67$                     & $0.67$                     \\
    DISTANCE             &  1.00  & 1  & $0.318$                    & $0.376$                    \\
    WORK\_INCOME\_TCR    & 46.35  & 2  & $8.60\times10^{-11}$$^{***}$& $1.12\times10^{-9}$$^{***}$\\
    ON\_DIALYSIS         &  9.88  & 1  & $0.0017$$^{**}$            & $0.0027$$^{**}$            \\
    ABO                  & 31.79  & 3  & $5.81\times10^{-7}$$^{***}$ & $1.89\times10^{-6}$$^{***}$\\
    PRA                  & 34.52  & 1  & $4.22\times10^{-9}$$^{***}$ & $2.74\times10^{-8}$$^{***}$\\
    AGE                  & 28.69  & 1  & $8.51\times10^{-8}$$^{***}$ & $3.69\times10^{-7}$$^{***}$\\
    GENDER               &  1.26  & 1  & $0.261$                    & $0.340$                    \\
    PREV\_KI\_TX         &  0.61  & 1  & $0.437$                    & $0.473$                    \\
    MED\_COND\_TRR       &  5.89  & 3  & $0.117$                    & $0.169$                    \\
    \bottomrule
  \end{tabular}
\end{table}

\subsubsection*{Stage 2: Association with Predictor Variable}

In the second stage, we tested whether each candidate variable $M_j \in M$ is statistically associated with the predictor $G$ (ethnicity). This step is essential for establishing the causal pathway required for mediation, as a variable can only mediate the relationship between ethnicity and waitlist duration if it is first affected by ethnicity.

We conducted likelihood ratio tests based on generalized linear or multinomial models appropriate to each mediator's structure. In all cases, we modeled the mediator as the response variable and ethnicity as the predictor, using the log-likelihood ratio statistic:
$$
\Lambda_j = -2 \left( \log L_{\text{null}} - \log L_{\text{full}} \right),
$$
which follows a chi-squared distribution with degrees of freedom equal to the number of parameters associated with $G$.

The specific model forms varied by mediator type:
\begin{itemize}
\item \textbf{Continuous mediators}: Linear regression models of the form $M_j = \gamma_0 + \gamma_G G + \varepsilon$ (full) versus $M_j = \gamma_0 + \varepsilon$ (null).
\item \textbf{Binary mediators}: Logistic regression with $\log\left( \frac{\mathbb{P}(M_j = 1)}{\mathbb{P}(M_j = 0)} \right) = \gamma_0 + \gamma_G G$ (full) versus $\log\left( \frac{\mathbb{P}(M_j = 1)}{\mathbb{P}(M_j = 0)} \right) = \gamma_0$ (null).
\item \textbf{Categorical mediators}: Multinomial logistic regression models comparing $\mathbb{P}(M_j = m \mid G = g) = \frac{\exp(\gamma_{0m} + \gamma_{gm})}{\sum_{m'} \exp(\gamma_{0m'} + \gamma_{gm'})}$ (full) against $\mathbb{P}(M_j = m) = \frac{\exp(\gamma_{0m})}{\sum_{m'} \exp(\gamma_{0m'})}$ (null).
\end{itemize}

\begin{table}[H]
  \centering
  \caption{Stage 2: Association Between Candidate Variables and Ethnicity}
  \label{tab:stage2_mediator_predictor}
  \setlength{\tabcolsep}{4pt}
  \renewcommand{\arraystretch}{1.1}
  \begin{tabular}{@{}lccc@{}}
    \toprule
    \textbf{Variable ($M_j$)} & \textbf{Test Type} & \textbf{p-value} & \textbf{BH-adjusted p-value} \\
    \midrule
    REGION               & MNLRT & $7.62\times10^{-149}$$^{***}$ & $9.90\times10^{-148}$$^{***}$ \\
    EDUCATION            & MNLRT & $1.77\times10^{-40}$$^{***}$  & $7.69\times10^{-40}$$^{***}$  \\
    CITIZENSHIP          & MNLRT & $5.84\times10^{-52}$$^{***}$  & $3.79\times10^{-51}$$^{***}$  \\
    PRI\_PAYMENT\_TCR\_KI & MNLRT & $2.54\times10^{-13}$$^{***}$  & $4.72\times10^{-13}$$^{***}$  \\
    DISTANCE             & LRT   & $0.0253$$^{*}$               & $0.0300$$^{*}$                \\
    WORK\_INCOME\_TCR    & MNLRT & $0.0011$$^{**}$              & $0.0016$$^{**}$               \\
    ON\_DIALYSIS         & LRT   & $4.93\times10^{-26}$$^{***}$ & $1.603\times10^{-25}$$^{***}$ \\
    ABO                  & MNLRT & $2.49\times10^{-21}$$^{***}$ & $6.48\times10^{-21}$$^{***}$  \\
    PRA                  & LRT   & $0.0899$                     & $0.0974$                      \\
    AGE                  & LRT   & $1.195\times10^{-14}$$^{***}$& $2.588\times10^{-14}$$^{***}$ \\
    GENDER               & LRT   & $0.0248$$^{*}$               & $0.0300$$^{*}$                \\
    PREV\_KI\_TX         & LRT   & $4.02\times10^{-5}$$^{***}$  & $6.54\times10^{-5}$$^{***}$   \\
    MED\_COND\_TRR       & MNLRT & $0.6150$                     & $0.6150$                      \\
    \bottomrule
  \end{tabular}
\end{table}

Note: LRT = Likelihood Ratio Test; MNLRT = Multinomial Likelihood Ratio Test.

\subsubsection*{Variable Classification}

Based on the BH-adjusted $p$-values from both stages, we classified variables according to the following criteria:

\begin{itemize}
\item \textbf{Mediators}: Variables significantly associated with both the outcome (Stage 1) and the predictor (Stage 2), with adjusted $p$-values $< 0.05$ in both stages.
\item \textbf{Covariates}: Variables significantly associated only with the outcome (Stage 1) but not with the predictor (Stage 2).
\item \textbf{Excluded}: Variables not significantly associated with the outcome in Stage 1.
\end{itemize}

Our structured testing procedure identified seven mediators (REGION, EDUCATION, CITIZENSHIP, WORK\_INCOME\_TCR, ON\_DIALYSIS, ABO, AGE) and one covariate (PRA) for inclusion in the Priority Fairness mediation analysis. Five variables were excluded because they showed insufficient association with the outcome variable. This classification ensures that our mediation model includes only variables with statistically supported relationships relevant to understanding how ethnicity influences waitlist duration.

\subsection*{S3.2 Variable Importance Analysis for Access Fairness Model}

To evaluate the factors influencing Living Kidney Donor Profile Index (LKDPI) scores in our Access Fairness analysis, we conducted a permutation-based variable importance analysis using a random forest model with 500 trees and 5-fold cross-validation. The model achieved strong predictive performance ($R^2 = 0.985$, MAE = 1.793) and included donor characteristics, transplant-specific factors, and recipient socioeconomic variables.

Permutation importance scores were computed by measuring the increase in prediction error when each variable's values are randomly permuted. The results are presented in Table~\ref{tab:variable_importance_access}.
\begin{table}[H]
\centering
\caption{Variable Importance Scores for LKDPI Prediction}
\label{tab:variable_importance_access}
\begin{tabular}{lc}
\hline
\textbf{Variable} & \textbf{Importance Score} \\
\hline
Donor Age & 100.00 \\
Donor Gender (Male) & 48.31 \\
Estimated GFR & 41.24 \\
Donor Black Ethnicity & 39.05 \\
Donor Smoking History & 35.38 \\
Systolic Blood Pressure & 31.06 \\
Recipient Gender (Male) & 27.02 \\
HLA-DR Mismatches & 14.89 \\
Donor-Recipient Weight Ratio & 12.34 \\
ABO Incompatibility & 11.60 \\
Donor BMI & 10.28 \\
HLA-B Mismatches & 5.75 \\
Recipient Ethnicity (Black) & 0.21 \\
Primary Payment (Private Insurance) & 0.14 \\
Transplant Year & 0.13 \\
Recipient Ethnicity (Hispanic) & 0.11 \\
Recipient Ethnicity (White) & 0.08 \\
\hline
\end{tabular}
\end{table}
The variable importance analysis demonstrates that donor characteristics are the primary determinants of LKDPI scores, with donor age, gender, and eGFR showing the highest importance. Critically, recipient ethnicity variables are of minimal importance (all < 0.25), indicating that recipient ethnic group membership has virtually no direct influence on kidney quality allocation, as measured by LKDPI. This finding supports the conclusion that observed disparities in kidney quality, if any, are not due to direct ethnic bias in the allocation system but rather may arise from systematic differences in the donor pools or matching processes across different patient populations.

\subsection*{S3.3 Variable Selection for Competing Risks Analysis}

To determine which covariates to include in our Outcome Fairness competing risks analysis, we implemented a systematic likelihood ratio testing procedure to evaluate the contribution of potential confounders to both graft rejection (Cause 1) and other causes of graft failure (Cause 2). This approach ensures that our model includes only variables with statistically supported associations with the outcomes of interest.

For each candidate variable, we conducted likelihood ratio tests comparing a full model including the variable to a reduced model excluding it. Using the competing risks framework with cause-specific Cox models, we separately tested each variable's contribution to both failure types. Let $h_k(t \mid G, W)$ represent the cause-specific hazard function for cause $k$ at time $t$, where $G$ denotes ethnicity and $W$ represents the vector of covariates. For each candidate variable $V_j \in W$, we compared a full model $h_k(t \mid G, W) = h_{0k}(t) \exp(\delta'G + \gamma'W)$ to a reduced model $h_k(t \mid G, W_{(-j)}) = h_{0k}(t) \exp(\delta'G + \gamma'W_{(-j)})$, where $W_{(-j)}$ represents the covariate vector excluding variable $V_j$. The likelihood ratio statistic $\text{LRT}_j = 2(\log L_{\text{full}} - \log L_{\text{reduced}})$ follows a $\chi^2$ distribution with degrees of freedom equal to the number of parameters associated with $V_j$.

We evaluated six candidate variables for inclusion in the competing risks model: ethnicity (primary variable of interest), UNOS transplant region, education level, citizenship status, post-transplant rejection treatment, and employment status. The results in Table~\ref{tab:competing_risks_lrt} show the likelihood ratio test statistics and p-values for each variable's contribution to both causes of graft failure.

\begin{table}[H]
\centering
\caption{Likelihood Ratio Test Results for Competing Risks Model Variable Selection. *p < 0.05, **p < 0.01, ***p < 0.001}
\label{tab:competing_risks_lrt}
\begin{tabular}{lcccc}
\hline
\multirow{2}{*}{\textbf{Variable}} & \multicolumn{2}{c}{\textbf{Cause 1 (Rejection)}} & \multicolumn{2}{c}{\textbf{Cause 2 (Other)}} \\
\cline{2-3} \cline{4-5}
& \textbf{LRT} & \textbf{p-value} & \textbf{LRT} & \textbf{p-value} \\
\hline
TRTREJ1Y\_KI & 36.92 & $9.59 \times 10^{-9}$*** & 54.62 & $1.38 \times 10^{-12}$*** \\
ETHCAT & 14.80 & 0.002** & 1.87 & 0.599 \\
WORK\_INCOME\_TRR & 0.15 & 0.926 & 37.60 & $6.84 \times 10^{-9}$*** \\
REGION & 11.74 & 0.303 & 16.96 & 0.075 \\
CITIZENSHIP & 3.28 & 0.351 & 6.39 & 0.094 \\
EDUCATION & 8.75 & 0.271 & 9.68 & 0.207 \\
\hline
\end{tabular}
\end{table}

Based on these results, we included variables that were statistically significant ($p < 0.05$) for at least one cause of graft failure. The final model includes ethnicity as the primary variable of interest (significant for rejection, $p = 0.002$), post-transplant rejection treatment (highly significant for both causes, $p < 0.001$), and employment status (significant for other causes, $p < 0.001$). We excluded education, region, and citizenship as they did not reach statistical significance for either cause, ensuring the model includes only the most clinically and statistically relevant confounders while avoiding overfitting.

Post-transplant rejection treatment showed the strongest association with both failure types, which is clinically expected as it directly relates to immunologic complications. Employment status demonstrated a significant association with non-rejection causes of graft failure, likely capturing socioeconomic factors that influence long-term transplant outcomes through medication adherence, healthcare access, and lifestyle factors. This systematic approach provides a balanced model that adequately controls for confounding while avoiding overfitting in the competing risks framework.

\end{document}